\documentclass{aa}  

\usepackage[dvipsnames]{xcolor}
\usepackage{graphicx}
\usepackage{array}
\usepackage{txfonts}

\usepackage{booktabs}
\usepackage{placeins}
\newcommand{\feI}{\ion{Fe}{i}}
\newcommand{\feII}{\ion{Fe}{ii}}
\newcommand{\snr}{S/N}
\newcommand{\planet}{WASP-76\,b}
\newcommand{\hoststar}{WASP-76}
\newcommand{\vsys}{v$_\mathrm{sys}$}
\newcommand{\kpvsys}{K$_\mathrm{p}$\,-\,v$_\mathrm{sys}$}
\newcommand{\kp}{K$_\mathrm{p}$}

\begin{document}

   \title{ESPRESSO reveals blueshifted neutral iron emission\\
   lines on the dayside of WASP-76\,b\,\thanks{Based on Guaranteed Time Observations collected at the European Southern Observatory under ESO programmes 1104.C-0350(U) and 110.24CD.004 by the ESPRESSO Consortium.}}


\author{
         \mbox{A. R. Costa Silva\inst{\ref{caup},\ref{dfa},\ref{geneve}}}
         \and
         \mbox{O. D. S. Demangeon\inst{\ref{caup},\ref{dfa}}}
         \and
         \mbox{N. C. Santos\inst{\ref{caup},\ref{dfa}}}
         \and
         \mbox{D. Ehrenreich\inst{\ref{geneve},\ref{cvu}}}
         \and
         \mbox{C. Lovis\inst{\ref{geneve}}}
         \and
         \mbox{H. Chakraborty\inst{\ref{geneve}}}
         \and
         \mbox{M. Lendl\inst{\ref{geneve}}}
         \and
         \mbox{F. Pepe\inst{\ref{geneve}}}
         \and
         \mbox{S. Cristiani\inst{\ref{inaf_trieste}}}
         \and
         \mbox{R. Rebolo\inst{\ref{iac},\ref{laguna},\ref{csic}}}
         \and
         \mbox{M. R. Zapatero-Osorio\inst{\ref{cab_madrid_mrzo}}}
         \and
         \mbox{V. Adibekyan\inst{\ref{caup},\ref{dfa}}}
         \and
         \mbox{Y. Alibert\inst{\ref{bern}}}
         \and
         \mbox{R. Allart\inst{\ref{montreal},\ref{geneve}}\thanks{Trottier Postdoctoral Fellow}}
         \and
         \mbox{C. Allende Prieto\inst{\ref{iac},\ref{laguna}}}
         \and
         \mbox{T. Azevedo Silva\inst{\ref{caup},\ref{dfa}}}
         \and
         \mbox{F. Borsa\inst{\ref{inaf_brera}}}
         \and
         \mbox{V. Bourrier\inst{\ref{geneve}}}
         \and
         \mbox{E. Cristo\inst{\ref{caup},\ref{dfa}}}
         \and
         \mbox{P. Di Marcantonio\inst{\ref{inaf_trieste}}}
         \and
         \mbox{E. Esparza-Borges\inst{\ref{iac},\ref{laguna}}}
         \and
         \mbox{P. Figueira\inst{\ref{geneve},\ref{caup}}}
         \and
         \mbox{J. I. Gonz\'alez Hern\'andez\inst{\ref{iac},\ref{laguna}}}
         \and
         \mbox{E. Herrero-Cisneros\inst{\ref{cab_madrid}}}
         \and
         \mbox{G. Lo Curto\inst{\ref{eso}}}
         \and
         \mbox{C. J. A. P. Martins\inst{\ref{caup},\ref{caup_solo}}}
         \and
         \mbox{A. Mehner\inst{\ref{eso}}} 
         \and
         \mbox{N. J. Nunes\inst{\ref{fcul}}}
         \and
         \mbox{E. Palle\inst{\ref{iac},\ref{laguna}}}
         \and
         \mbox{S. Pelletier\inst{\ref{geneve}}}
         \and
         \mbox{J. V. Seidel\inst{\ref{eso}}}
         \and
         \mbox{A. M. Silva\inst{\ref{caup},\ref{dfa}}}
         \and
         \mbox{S. G. Sousa\inst{\ref{caup}}}
         \and 
         \mbox{A. Sozzetti\inst{\ref{inaf_torino}}}
         \and
         \mbox{M. Steiner\inst{\ref{geneve}}}      
         \and
         \mbox{A. Su{\'a}rez Mascare{\~n}o\inst{\ref{iac},\ref{laguna}}}
         \and
         \mbox{S. Udry\inst{\ref{geneve}}}
         }

   \institute{
             Instituto de Astrof\'isica e Ci\^encias do Espa\c{c}o, Universidade do Porto, CAUP, Rua das Estrelas, 4150-762 Porto, Portugal \label{caup}
             \and
             Departamento de F\'isica e Astronomia, Faculdade de Ci\^encias, Universidade do Porto, Rua do Campo Alegre, 4169-007 Porto, Portugal \label{dfa}
             \and
             Observatoire Astronomique de l'Universit\'e de Gen\`eve, Chemin Pegasi 51, 1290 Versoix, Switzerland \label{geneve} 
             \and
             Centre Vie dans l'Univers, Facult\'e des sciences, Universit\'e de Gen\`eve, Gen\`eve 4, Switzerland \label{cvu}
             \and
             INAF $-$ Osservatorio Astronomico di Trieste, via G. B. Tiepolo 11, I-34143, Trieste, Italy \label{inaf_trieste} 
             \and
             Instituto de Astrof{\'\i}sica de Canarias, c/ V\'ia L\'actea s/n, 38205 La Laguna, Tenerife, Spain \label{iac} 
             \and
             Departamento de Astrof\'{\i}sica, Universidad de La Laguna, 38206 La Laguna, Tenerife, Spain \label{laguna} 
             \and
             Consejo Superior de Investigaciones Cient\'{\i}cas, Spain \label{csic}
             \and
             Centro de Astrobiolog\'ia, CSIC-INTA, Camino Bajo del Castillo s/n, E-28692 Villanueva de la Ca\~nada, Madrid, Spain \label{cab_madrid_mrzo}
             \and
             Physics Institute, University of Bern, Sidlerstrasse 5, 3012 Bern, Switzerland \label{bern}
             \and
             D\'epartement de Physique, Institut Trottier de Recherche sur les Exoplan\`etes, Universit\'e de Montr\'eal, Montr\'eal, Qu\'ebec, H3T 1J4, Canada \label{montreal}
             \and
             INAF $-$ Osservatorio Astronomico di Brera, Via E. Bianchi 46, 23807 Merate (LC), Italy \label{inaf_brera}
             \and
             Centro de Astrobiolog\'ia, CSIC-INTA, Crta. Ajalvir km 4, E-28850 Torrej\'on de Ardoz, Madrid, Spain \label{cab_madrid}
             \and
             European Southern Observatory, Alonso de C\'ordova 3107, Vitacura, Regi\'on Metropolitana, Chile \label{eso} 
             \and
             Centro de Astrof\'{\i}sica da Universidade do Porto, Rua das Estrelas, 4150-762 Porto, Portugal \label{caup_solo}
             \and
             Instituto de Astrof\'isica e Ci\^encias do Espa\c{c}o, Faculdade de Ci\^encias da Universidade de Lisboa, Campo Grande, PT1749-016 Lisboa, Portugal \label{fcul}
             \and
             INAF $-$ Osservatorio Astrofisico di Torino, Via Osservatorio 20, 10025 Pino Torinese, Italy \label{inaf_torino}
            }
             
     \titlerunning{ESPRESSO reveals blueshifted neutral iron emission lines on the dayside of WASP-76\,b}
     \authorrunning{Costa Silva et al. (2024)}

   \date{Received 11 March 2024; accepted 9 July 2024}

 
  \abstract
   {Ultra hot Jupiters (gas giants with T$_\mathrm{eq}>2000$\,K) are intriguing exoplanets due to the extreme physics and chemistry present in their atmospheres. Their torrid daysides can be characterised using ground-based high-resolution emission spectroscopy.}
   {We search for signatures of neutral and singly ionised iron (\feI\ and \feII, respectively) in the dayside of the ultra hot Jupiter WASP-76\,b, as these species were detected via transmission spectroscopy in this exoplanet. Furthermore, we aim to confirm the existence of a thermal inversion layer, which has been reported in previous studies, and attempt to constrain its properties.}
   {We observed WASP-76\,b on four epochs with ESPRESSO at the VLT, at orbital phases shortly before and after the secondary transit, when the dayside is in view. We present the first analysis of high-resolution optical emission spectra for this exoplanet. We compare the data to synthetic templates created with \texttt{petitRADTRANS}, using cross-correlation function techniques.}
   {We detect a blueshifted ($-$4.7$\,\pm\,$0.3\,km/s) \feI\ emission signature on the dayside of WASP-76\,b at 6.0\,$\sigma$. The signal is detected independently both before and after the eclipse, and it is blueshifted in both cases. The presence of iron emission features confirms the existence of a thermal inversion layer. \feII\ was not detected, possibly because this species is located in the upper layers of the atmosphere, which are more optically thin. Thus the \feII\ signature on the dayside of \planet\ is too weak to be detected with emission spectroscopy.}
   {We propose that the blueshifted \feI\ signature is created by material rising from the hot spot to the upper layers of the atmosphere, and discuss possible scenarios related to the position of the hotspot. This work unveils some of the dynamic processes ongoing on the dayside of the ultra hot Jupiter WASP-76\,b through the analysis of the \ion{Fe}{i} signature from its atmosphere, and complements previous knowledge obtained from transmission studies. It also highlights the ability of ESPRESSO to probe the dayside of this class of exoplanets.}


\maketitle
%

\section{Introduction} \label{intro}

\begin{table*}[h]
    \caption{Properties of the planet \planet\ and its stellar host.}
    \centering
    \begin{tabular}{lccc}
        \hline \hline
        Parameter & Symbol (unit) & Value & Reference \\
        \hline
        \noalign{\smallskip}
        \textbf{Star} &&& \\
        \noalign{\smallskip}
        Right ascension         & RA                                &  01\,h 46\,m 31.9\,s          & \citet{gaia2020} \\
        Declination             & DEC                               & +02$^{\circ}$ 42' 02.0''         & \citet{gaia2020} \\
        V magnitude             & V-mag (mag)                              & 9.52\,$\pm$\,0.03             & \citet{hog2000} \\
        Spectral type           &                                   & F7                            & \citet{west2016} \\
        Effective temperature   & T$_\mathrm{eff}$ (K)              & 6329\,$\pm$\,65               & \citet{ehrenreich2020} \\
        Radius                  & R$_*$ (R$_\odot$)                 & 1.458\,$\pm$\,0.021           & \citet{ehrenreich2020} \\
        Mass                    & M$_*$ (M$_\odot$)                 & 1.756\,$\pm$\,0.071           & \citet{ehrenreich2020} \\
        Gravity                 & log\,$g$                          & 4.196\,$\pm$\,0.106           & \citet{ehrenreich2020} \\
        RV semi-amplitude       & K$_*$ (m/s)                       & 116.02\,$^{+1.29}_{-1.35}$    & \citet{ehrenreich2020} \\
        Systemic velocity       & v$_\mathrm{sys}$ (km/s)           &  &  \\
        \hspace{1cm} Epoch I    &  & $-$1.2113\,$\pm$0.0003              & this work \\
        \hspace{1cm} Epoch II   &  & $-$1.2134\,$\pm$0.0003              & this work \\
        \hspace{1cm} Epoch III  &  & $-$1.2064\,$\pm$0.0002              & this work \\
        \hspace{1cm} Epoch IV   &  & $-$1.2100\,$\pm$0.0003              & this work \\
        \hline
        \noalign{\smallskip}
        \textbf{Planet} &&& \\
        \noalign{\smallskip}
        Radius                  & R$_\mathrm{p}$ (R$_\mathrm{Jup}$) & 1.854\,$^{+0.077}_{-0.076}$   & \citet{ehrenreich2020} \\ 
        Mass                    & M$_\mathrm{p}$ (M$_\mathrm{Jup}$) & 0.894\,$^{+0.014}_{-0.013}$   & \citet{ehrenreich2020} \\
        Planet surface gravity  & g$_\mathrm{p}$ (m/s$^2$)                   & 6.4\,$\pm$\,0.5               & \citet{ehrenreich2020} \\
        Inclination             & i (deg)                         & 89.623\,$^{+0.014}_{-0.013}$  & \citet{ehrenreich2020} \\
        Semi-major axis         & a (AU)                          & 0.0330\,$\pm$\,0.0002         & \citet{ehrenreich2020} \\
        Orbital period          & P (days)                        & 1.80988198\,$^{+0.00000064}_{0.00000056}$ & \citet{ehrenreich2020} \\
        Mid-transit time        & T$_\mathrm{c}$ (BJD)              & 58080.626165\,$^{+0.000418}_{0.000367}$    & \citet{ehrenreich2020} \\
        RV semi-amplitude       & \kp\ (km/s)             & 196.52\,$\pm$\,0.94           & \citet{ehrenreich2020} \\
        Equilibrium temperature & T$_\mathrm{eq}$ (K)               & 2228\,$\pm$\,122              & \citet{ehrenreich2020} \\
        Dayside temperature     & T$_\mathrm{day}$ (K)              & 2693\,$\pm$\,56               & \citet{garhart2020} \\
         \hline
    \end{tabular}
    \label{tab:properties}
\end{table*}

\begin{table}[h!]
    \caption{Literature reports of atom, ion, and molecule detections in the atmosphere of \planet.}
    \centering
    \begin{tabular}{lc}
         \hline \hline
         Chemical species & Reference \\
         \hline
         \ion{Ba}{ii} & [1], [2] \\
         \ion{Ca}{i} & [2]\\
         \ion{Ca}{ii} & [1], [2] ,[3], [4], [5], [6], [7] \\
         \ion{Co}{i} & [6] \\
         \ion{Cr}{i} & [1], [2], [6], [7] \\
         \ion{Fe}{i} & [1], [2], [3], [6], [7], [8], [9] \\
         \ion{Fe}{ii} & [2] \\
         \ion{H}{i} & [1], [2], [6] \\
         \ion{He}{i} & [4], [10] \\
         \ion{K}{i} & [2], [3], [6], [7] \\
         \ion{Li}{i} & [1], [2], [3], [6], [7] \\
         \ion{Mg}{i} & [1], [2], [3], [6] \\
         \ion{Mn}{i} & [1], [2], [3], [6] \\
         \ion{Na}{i} & [1], [2], [3], [5], [6], [7],\\
                     & [11], [12], [13], [14] \\
         \ion{Ni}{i} & [2], [6] \\
         \ion{O}{i} & [2] \\
         \ion{Sr}{ii} & [6] \\
         \ion{V}{i} & [1], [2], [6], [7] \\
         \noalign{\smallskip} \hline \noalign{\smallskip}
         CO & [15], [16] \\
         H$_\mathrm{2}$O & [15], [16], [17], [18], [19], [20] \\
         HCN & [17] \\
         OH & [21] \\
         TiO & [15], [18], [22] \\
         VO & [2], [18], [20] \\
          \hline
    \end{tabular}
    \tablebib{(1)~\citet{azevedosilva2022};
        (2) \citet{pelletier2023}; (3) \citet{tabernero2021}; (4) \citet{casasayas-barris2021}; (5) \citet{deibert2021}; (6) \citet{kesseli2022}; (7) \citet{deibert2023}; (8) \citet{ehrenreich2020}; (9) \citet{kesseli2021}; (10) \citet{lampon2023}; (11) \citet{seidel2019}, (12) \citet{zak2019}, (13) \citet{seidel2021}, (14) \citet{kawauchi2022}, (15) \citet{fu2021}, (16) \citet{yan2023}, (17) \citet{sanchez-lopez2022}, (18) \citet{tsiaras2018}, (19) \citet{fisherheng2018}, (20) \citet{changeat2022}, (21) \citet{landman2021}, (22) \citet{edwards2020}.}
    \label{tab:w76_detections}
\end{table}

In recent years, the field of exoplanet research has partially shifted its focus from the detection of other worlds to their detailed characterisation. The study of exoplanet atmospheres provides extensive knowledge that helps us constrain the formation and evolution models of exoplanetary systems \citep[e.g.][]{madhusudhan2019}, and it is also considered a crucial step in the search for extraterrestrial life \citep[e.g.][]{schwieterman2018, meadows2018}.

Ultra hot Jupiters (UHJs) are an interesting class of exoplanets to further characterise, due to their extreme equilibrium temperatures (T$_\mathrm{eq}>$\,2000\,K) and even hotter dayside temperatures. It has been predicted that thermal inversion layers exist on the dayside of these scorched planets \citep[e.g.][]{lothringer2018, lothringer2019}, that is a hotter layer of atmosphere on top of a colder layer, where chemical species produce emission features. Recent observational studies have confirmed this by probing the emission spectra of exoplanets during orbital phases close to the secondary transit (also known as eclipse or occultation), and detecting such features \citep[e.g.][]{evans2017, yan2020, borsa2022, changeat2022}.

However, it is still unclear what species are absorbing the stellar irradiation at high altitudes and creating the inversion. For some time, titanium oxide (TiO) and vanadium oxide (VO) were assumed to be the opacity sources responsible for this feature \citep{hubeny2003, fortney2008}. Though more recently, \citet{lothringer2018} have shown that neutral iron (\feI) and other atomic metals are also capable of creating an inversion in the atmospheres of UHJs, without the need for TiO or VO. \feI\ detections have been reported in multiple UHJs, both in the terminators and on the dayside, through transmission spectroscopy \citep[e.g.][]{hoeijmakers2018, ehrenreich2020, bourrier2020, borsa2021} and emission spectroscopy \citep[e.g.][]{pino2020, scandariato2023}, respectively. 

Detecting singly ionised iron (\feII) is also of interest, as it could shed some light on the effect that magnetic fields have on the atmosphere of UHJs \citep[e.g.][]{perna2010}, and inform modelling works regarding the ionisation (and recombination) fraction of iron. \feII\ has been detected in several hot giants via transmission studies \citep[e.g.][]{hoeijmakers2019, borsa2021, prinoth2022, belloarufe2022}, however, it has only been detected once in emission observations of a UHJ. \citet{borsa2021} reported \feII\ in the post-eclipse observation of KELT-20\,b/MASCARA-2\,b, but follow-up studies have not been able to confirm the detection \citep{yan2022, kasper2023, petz2024}. Further non-detections have been reported for KELT-9\,b \citep{pino2020, riddenharper2023}, WASP-33\,b \citep{cont2022}, and WASP-121\,b \citep{hoeijmakers2024}.

The target of our study is \planet\ \citep{west2016}, an UHJ orbiting an F7-star (V\,=\,9.52\,mag) on a 1.81-day period, at a separation of 0.033 AU (the parameters of \planet\ and its host star are listed in Table \ref{tab:properties}). The equilibrium temperature of this exoplanet is T$_\mathrm{eq}\,\sim\,2228$\,K \citep{ehrenreich2020}, but as it is tidally locked, the dayside can reach temperatures up to T$_\mathrm{dayside}\,\sim\,$2693\,K \citep{garhart2020}. At a radius of 1.83\,R$_\mathrm{Jup}$ and a mass of 0.92\,M$_\mathrm{Jup}$, \planet\ is a benchmark UHJ that has been frequently investigated in recent years. Several atomic, ionised and molecular species have been detected in the atmosphere of \planet, which are summarised in Table \ref{tab:w76_detections}. Furthermore, this exoplanet has been the target of phase curve observations \citep{tsiaras2018, garhart2020, may2021, fu2021, demangeon2024} and global circulation models (GCMs) have been developed to explain the observed signatures and better understand the mechanics of this atmosphere \citep[e.g.][]{may2021, savel2022, wardenier2021, wardenier2023, schneider2022, beltz2022a, beltz2022b, beltz2023, sainsbury-martinez2023, demangeon2024}.

In particular, the detection of \feI\ by \citet{ehrenreich2020} motivates our work. The authors first reported an asymmetrical \feI\ detection in the terminators of \planet, when analysing high-resolution transmission spectra from ESPRESSO\footnote{Echelle SPectrograph for Rocky Exoplanets and Stable Spectroscopic Observations} \citep{pepe2021}. This finding was supported by \citet{kesseli2021, kesseli2022}, and \citet{pelletier2023}, who independently recovered the asymmetrical signature. To explain why \feI\ is observed on the evening limb but not on the morning limb, \citet{ehrenreich2020} suggest that this species must condense as it crosses the colder nightside, and thus should exist in gaseous form on the dayside of \planet. Furthermore, \citet{pelletier2023} also report a tentative detection of \feII\ in transmission studies of the atmosphere. Thus we aim to determine whether iron, in both the neutral and ionised forms, is present on the dayside of this planet via emission spectroscopy.

To the extent of our knowledge, only one high-resolution emission study has been published for this target so far. \citet{yan2023} analysed data from CRIRES+\footnote{The CRyogenic InfraRed Echelle Spectrograph Upgrade Project} \citep{dorn2023} and reported the detection of CO and weak H$_2$O emission features. The CO signature was slightly redshifted (2.1$^{+0.8}_{-0.7}$\,km/s), whereas the H$_2$O signal had a small blueshift ($-$5$^{+12}_{-10}$\,km/s, though we note the large uncertainty). These emission lines confirmed the existence of a thermal inversion on the dayside of \planet, as had been suggested from previous low-resolution spaceborne observations \citep{edwards2020, may2021}.

High-resolution spectroscopy has been proven fundamental in the endeavour of characterising atmospheres, as it allows us to distinguish a forest of individual lines in the spectra. Additionally, due to the different Doppler shifts of the stellar and planetary signal over time, we are often able to disentangle the stellar contribution from the planetary features in high-resolution data \citep{snellen2010}. In some cases, the line strength is large enough to allow for the analysis of lines individually \citep{wyttenbach2015, allart2018, seidel2019, allart2023}. However, for weaker spectral features, we can instead employ a cross-correlation function (CCF) to harvest the signal of multiple lines at once, instead of individually \citep[e.g.][]{snellen2010, brogi2012, birkby2013, hoeijmakers2015, allart2017, ehrenreich2020, azevedosilva2022, prinoth2023}. Since all the planetary lines are shifted by the same radial velocity (RV) value, the robustness of the CCF is enhanced, and so is the significance of the detections.

In this work, we present the first analysis of high-resolution emission spectra at visible wavelengths for \planet\ obtained with ESPRESSO at the Very Large Telescope (VLT). The main goal is to probe the existence of \feI\ and \feII\ in the exoplanet's dayside, which could be contributing to the inverted pressure-temperature (P-T) profile. The thermal inversion layer has been previously confirmed in \citet{yan2023} from H$_2$O emission lines, but observing the emission features of iron can help further constrain the thermal structure of the planet's dayside.

This paper is structured as follows: in Sect. \ref{obs}, we describe the details of our observations; in Sect. \ref{analysis}, we present the data reduction, processing, and analysis steps for computing the CCFs; in Sect. \ref{results}, we report the results of the analysis; we compare our findings to the literature in Sect. \ref{discussion} and propose an atmospheric scenario that could link our results with those from the literature; finally, we summarise and conclude in Sect. \ref{summary}.

\section{Observations} \label{obs}

\begin{figure*}[h]
    \centering
    \includegraphics[width=0.5\hsize]{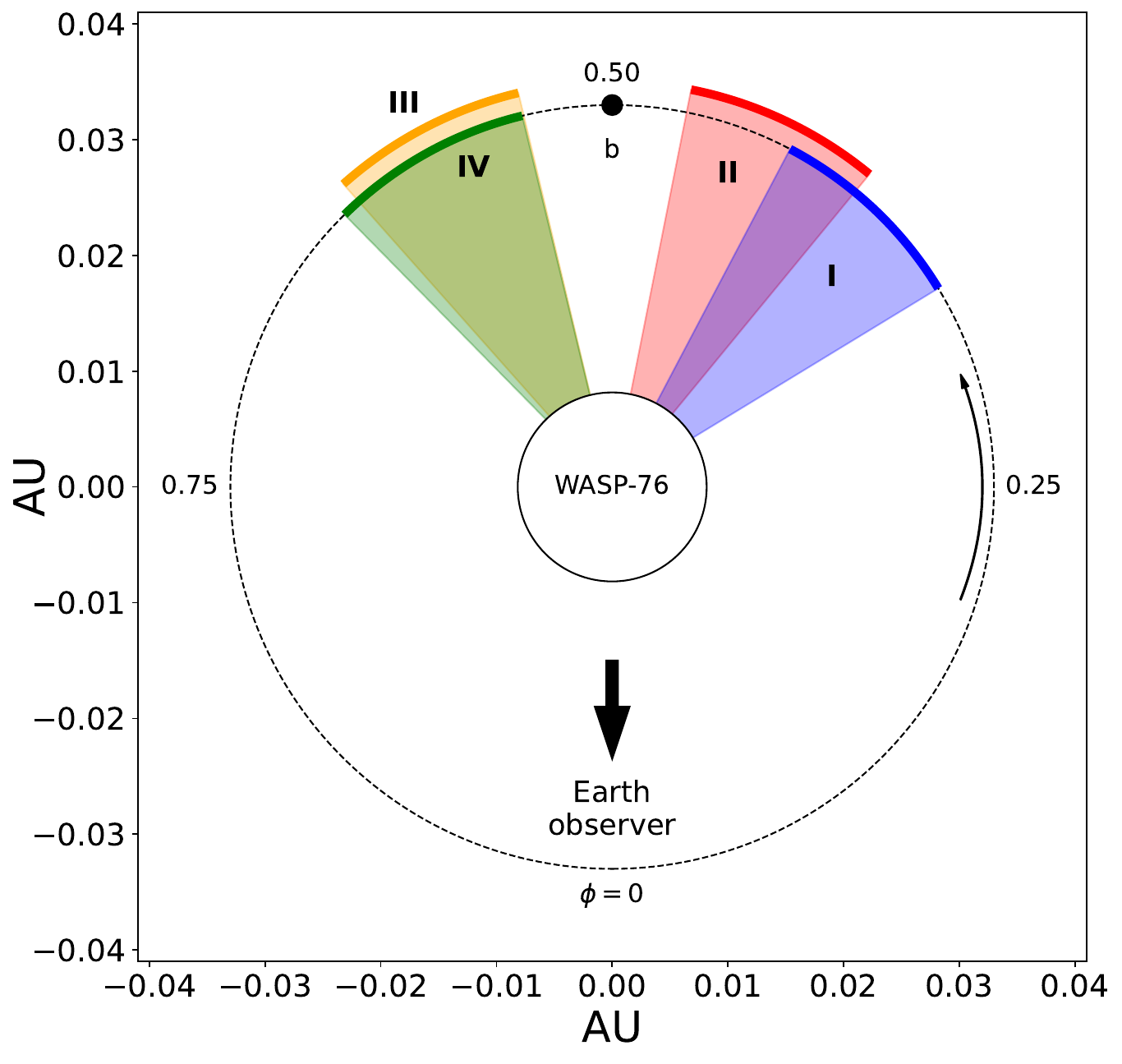}
    \includegraphics[width=0.44\hsize]{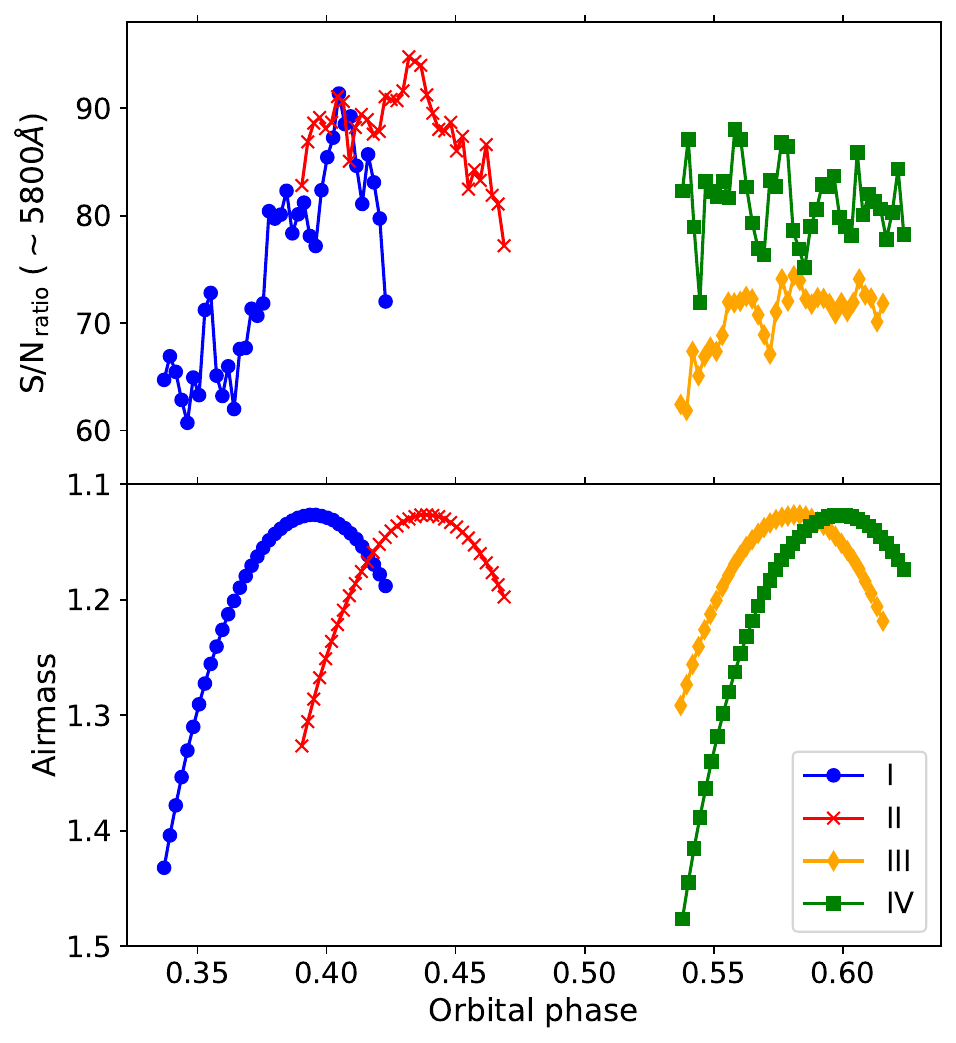}
    \caption{Details of the observations of WASP-76\,b with ESPRESSO. \textbf{Left:} Orbital diagram of \planet\ showing the epochs and phases ($\phi$) during which the system was observed. The curved arrow indicates increasing orbital phase for the planet. \textbf{Right:} Variation of signal-to-noise ratio around 5800$\,\AA$ (\textit{top}) and airmass (\textit{bottom}) for each epoch observed. See more details in Table \ref{tab:obs_log}.}
    \label{fig:observations}
\end{figure*}

\begin{table*}[h]
    \caption{Observation logs.}
    \label{tab:obs_log}
    \centering
    \begin{tabular}{lccccccc}
        \hline \hline
        Data set & Night & VLT Unit & \# spectra & Airmass range & Exposure time & Phase coverage & \snr\ range$^\ast\,$ \\
        \hline
        I   & 2022-10-14 & UT1 & 39 & 1.13\,$-$\,1.43 & 300 s & 0.34\,$-$\,0.42 & 61\,$-$\,91  \\
        II  & 2021-09-09 & UT2 & 35 & 1.13\,$-$\,1.33 & 300 s & 0.39\,$-$\,0.47 & 77\,$-$\,94  \\
        III & 2021-09-02 & UT1 & 35 & 1.13\,$-$\,1.29 & 300 s & 0.54\,$-$\,0.61 & 62\,$-$\,74  \\ 
        IV  & 2022-10-18 & UT1 & 39 & 1.13\,$-$\,1.48 & 300 s & 0.54\,$-$\,0.62 & 72\,$-$\,88  \\
        \hline
    \end{tabular}
    \tablefoot{$^\ast\,$Values taken from order 106 of ESPRESSO, at $\sim5800\AA$.}
\end{table*}

\begin{figure*}
\sidecaption
  \includegraphics[width=12cm]{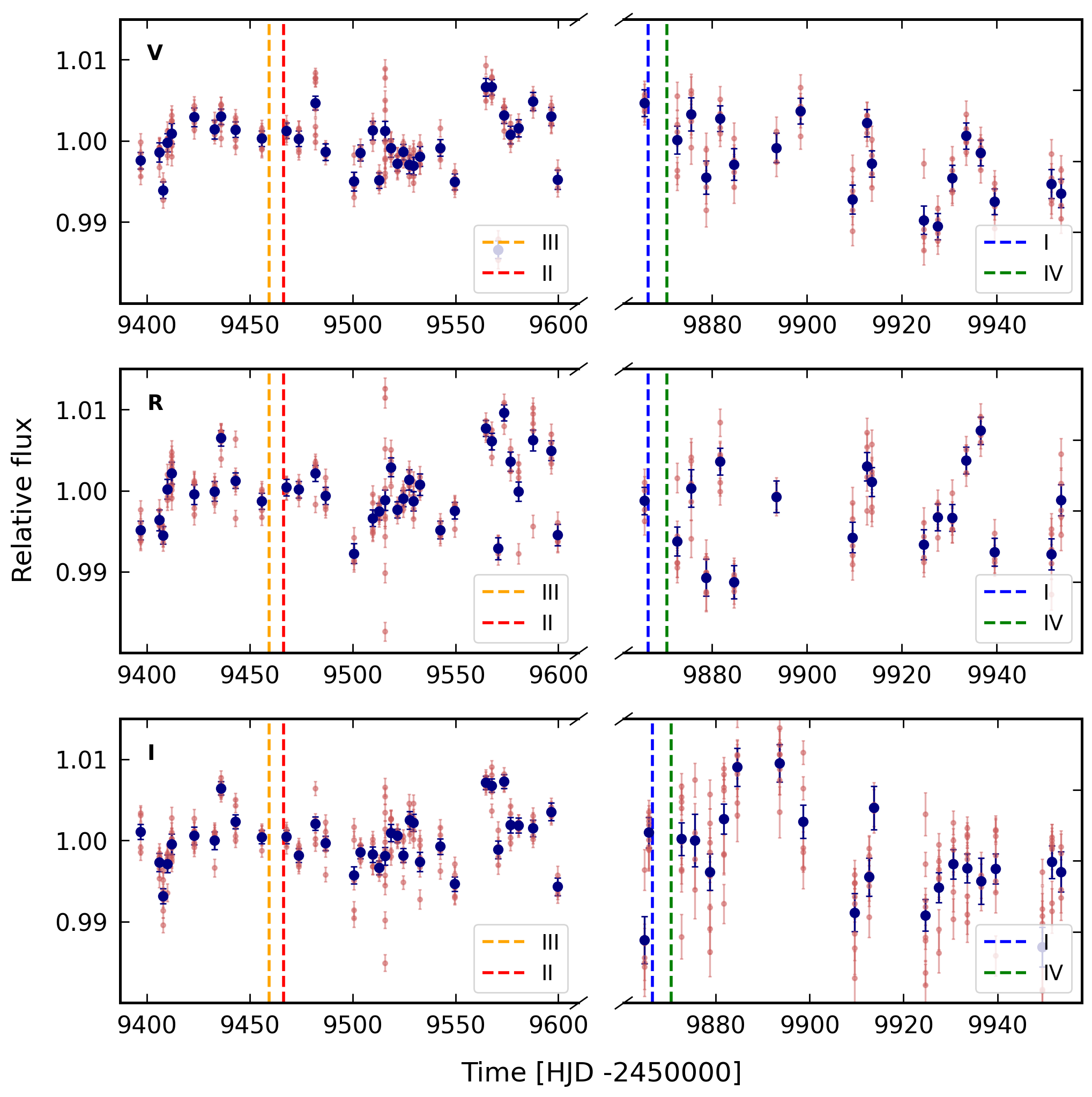}
     \caption{Photometric monitoring of WASP-76: EulerCam light curves in Johnson-$V$ (top), Sloan-$r'$ (middle), and Sloan-$i'$ (bottom) filters, unbinned (red) and binned per day (blue). WASP-76 can be considered photometrically quiet around the time of the ESPRESSO observations (vertical dashed lines).}
     \label{fig:multi-lcs}
\end{figure*}

\subsection{ESPRESSO spectroscopic data}


The WASP-76 system was observed on four different epochs, two of them covering phases before the planet's secondary transit (2021 September 9 and 2022 October 14) and two after (2021 September 2 and 2022 October 18). The data were obtained with the ESPRESSO spectrograph (Pepe et al. 2021) installed at the VLT, at Cerro Paranal, as part of the programmes 1104.C-0350(U) and 110.24CD.004 of the Guaranteed Time Observation. The ESPRESSO mode was set to HR21 (1 unit telescope, 2\,$\times$\,1 binning), with resolving power of R\,$\sim$\,140\,000. Fibre A was pointed to our target whereas Fibre B was pointing towards the sky to allow for sky subtraction. Each spectrum covers the visible wavelength range from $\sim$3800\,$\AA$ to $\sim$7880\,$\AA$. 

This resulted in a total of 148 high-resolution spectra, evenly split between the pre- and post-eclipse observations (74 spectra probing each phase range). The observations lasted between 3.5$\,$h and 3.75$\,$h for each epoch, with the individual exposure time set to 300 seconds. The left panel of Fig. \ref{fig:observations} illustrates the orbital phases covered by the observations. We label each epoch as data sets I, II, III, and IV, in order of increasing orbital phase at the start of observations. The right panels of this figure show the variation of airmass in each epoch, as well as the signal-to-noise ratio (\snr) of each data set (taken from order 106 of ESPRESSO, at wavelength $\sim\,$5800$\,\AA$). The observation logs are summarised in Table \ref{tab:obs_log}.

\subsection{EulerCam photometric data}

We searched for potential photometric variability around the time of the ESPRESSO observations to rule out the presence of strong active regions on the star that could contaminate the retrieved emission spectrum. For this, we observed WASP-76 with EulerCam (\citealp{lendl_2012}), a 4k\,$\times$\,4k CCD detector installed at the Cassegrain focus of the 1.2-m \textit{Leonhard Euler} Telescope at ESO's La Silla Observatory. We monitored the star every third night in two different monitoring campaigns: i) from 2021 June 20 to 2022 January 20, and ii) from 2022 October 1 to 2023 January 10. Each monitoring observation involved taking a sequence of six images in five different filters, with exposure times indicated in parenthesis: Johnson-$B$ (60 s), Johnson-$V$ (30 s), Sloan-$r'$ (20 s), Sloan-$i'$ (30 s) and Sloan-$z'$ (40 s).


The raw full-frame EulerCam images for each night of observation are corrected for over-scan, bias and flat-field using the standard reduction pipeline \citep{lendl_2012}. The aperture photometry is performed using circular apertures with radii ranging from 16 to 80 pixels, placed on the target star and three bright stars in the field of view. To mitigate the changing position of stars on the detector, the placement of apertures was performed using their astrometric solution \citep{dustin_lang_2010}. The optimal aperture for each night is found by minimising the photometric scatter of the consecutive images. The light curves from Johnson-$B$ and Sloan-$z'$ were removed as they suffered from strong systematics. The normalised differential light curves for Johnson-$V$, Sloan-$r'$, and Sloan-$i'$ are shown in Fig. \ref{fig:multi-lcs}. 

WASP-76 appears to be photometrically quiet around the ESPRESSO eclipse observations, with upper limits on the flux variability from the light curves being 4.95 mmag ($V$), 7.17 mmag ($r'$), and 5.55 mmag ($i'$). We note that our observations cannot resolve the two components of the WASP-76 binary system \citep{wollert2015}. The companion star, WASP-76\,B, is likely a late-G or early K-type dwarf \citep{ehrenreich2020}, and it lies at a separation of $\sim\,$0.44\,\arcsec, which corresponds to 53.0\,$\pm$\,8.8\,AU \citep{ginski2016, ngo2016, bohn2020}. Though the stars are unresolved, our analysis indicates no photometric activity in general, which would suggest both components A and B are quiet.

\section{Analysis} \label{analysis}

\subsection{Reducing ESPRESSO data and extracting planet's spectra}

The raw data were reduced with the ESPRESSO Data Reduction System \citep[DRS, version 3.0.0,][]{pepe2021}. We proceeded to analyse the S1D sky-subtracted spectra produced by this pipeline, which is in the rest frame of the barycenter of the Solar System. In S1D, all orders of the spectrograph have been merged into a single 1D spectrum for each exposure. Our analysis follows a similar procedure to previous works, and its seven steps are detailed below:

\begin{itemize}
    \item Remove telluric contamination.\\
    We removed the telluric contamination using Molecfit \citep[version 4.2,][]{smette2015, kausch2015} with the ESPRESSO settings. In some regions of the spectra, the telluric features are completely saturated, which makes it impossible to apply a correction. In the following wavelength ranges (in air), we could not achieve a satisfactory correction and thus they are masked at later stages when calculating the CCF: [5867.56 $-$ 6005.55]\,$\AA$, [6270.23 $-$ 6344.15]\,$\AA$, [6439.08 $-$ 6606.96]\,$\AA$, [6858.15 $-$ 7417.40]\,$\AA$, [7586.03 $-$ 7751.12]\,$\AA$.

    \item Normalise spectra.\\
    The continuum contribution was removed from the spectra with RASSINE \citep{cretignier2020}, via the S-BART Python package \citep{silva2022}, which optimises the process for ESPRESSO data. This step removes the interference patterns that have been reported to affect the continuum of ESPRESSO observations (commonly referred to as wiggles).

    \item Fit for the systemic velocity.\\
    We found that there was a discrepancy in the systemic velocity (\vsys) values presented in the literature. The discovery paper reports this value as $-$1.0733$\,\pm\,$0.0002\,km/s \citep{west2016}; SIMBAD gives $-$1.152$\,\pm\,$0.0033\,km/s, from the Gaia Data Release 2 \citep{soubiran2018}; and the analysis of \citet{ehrenreich2020} states different values for each epoch observed: $-$1.162, $-$1.167, and $-$1.171\,km/s (with typical uncertainty of the order of 0.002\,km/s). This value warrants attention as it is an important parameter to accurately shift the spectra between the different rest frames. Thus we chose to perform a simple least-squares fit of a Keplerian to the RV values calculated by the pipeline, setting \vsys\ as the only free parameter (the remaining orbital elements were set to those reported in Table \ref{tab:properties}). We obtained four values of \vsys: $-$1.2113, $-$1.2134, $-$1.2064, $-$1.2100\,km/s, for epochs I, II, III, and IV, respectively. The typical uncertainty obtained from this fit is of the order of 0.0003\,km/s, though the real uncertainty on \vsys\ is expected to be much greater. In the following steps, the data of each epoch was processed using the corresponding \vsys\ retrieved here.

    \item Create stellar template.\\
    All of the spectra were shifted to the stellar rest frame and the median spectrum was computed for each epoch independently. The median spectra contain only the stellar lines, which are aligned at the same position in the star's rest frame, increasing their signal. The planetary lines are very faint and their position on the spectra changes significantly over time (the planet's RV changes by $\sim$\,2.4\,km/s between the start of two consecutive exposures). Thus their contribution is diluted when computing the median template in the stellar rest frame, averaging out to values comparable to the noise.

    \item Extract planet's spectra.\\
    To obtain spectra that contain only the planetary features, we interpolate all spectra to a common wavelength grid and subtract the stellar median template from every spectrum of the corresponding epoch, in the stellar rest frame.

    \item Compute cross-correlation.\\
    We compare each exposure of the planetary spectra with synthetic models created with \texttt{petitRADTRANS} \citep[][see next section]{molliere2019, molliere2020, alei2022}, by computing a non-weighted CCF according to:

\begin{equation}
    \label{eq:ccf}
    \begin{centering}
        \mathrm{CCF\,(RV)} = \sum_i s_i m_{i}\mathrm{(RV)}
    \end{centering}
\end{equation}

    where $s_i$ is each data point in the planet's spectrum, and $m_{i}\mathrm{(RV)}$ is each data point of the model shifted by a given RV lag. At this stage, each spectrum produces one CCF curve, with velocities ranging from -300 to 300\,km/s, with a step of 1\,km/s.


\begin{figure*}[h!]
    \centering
    \includegraphics[width=0.37\hsize]{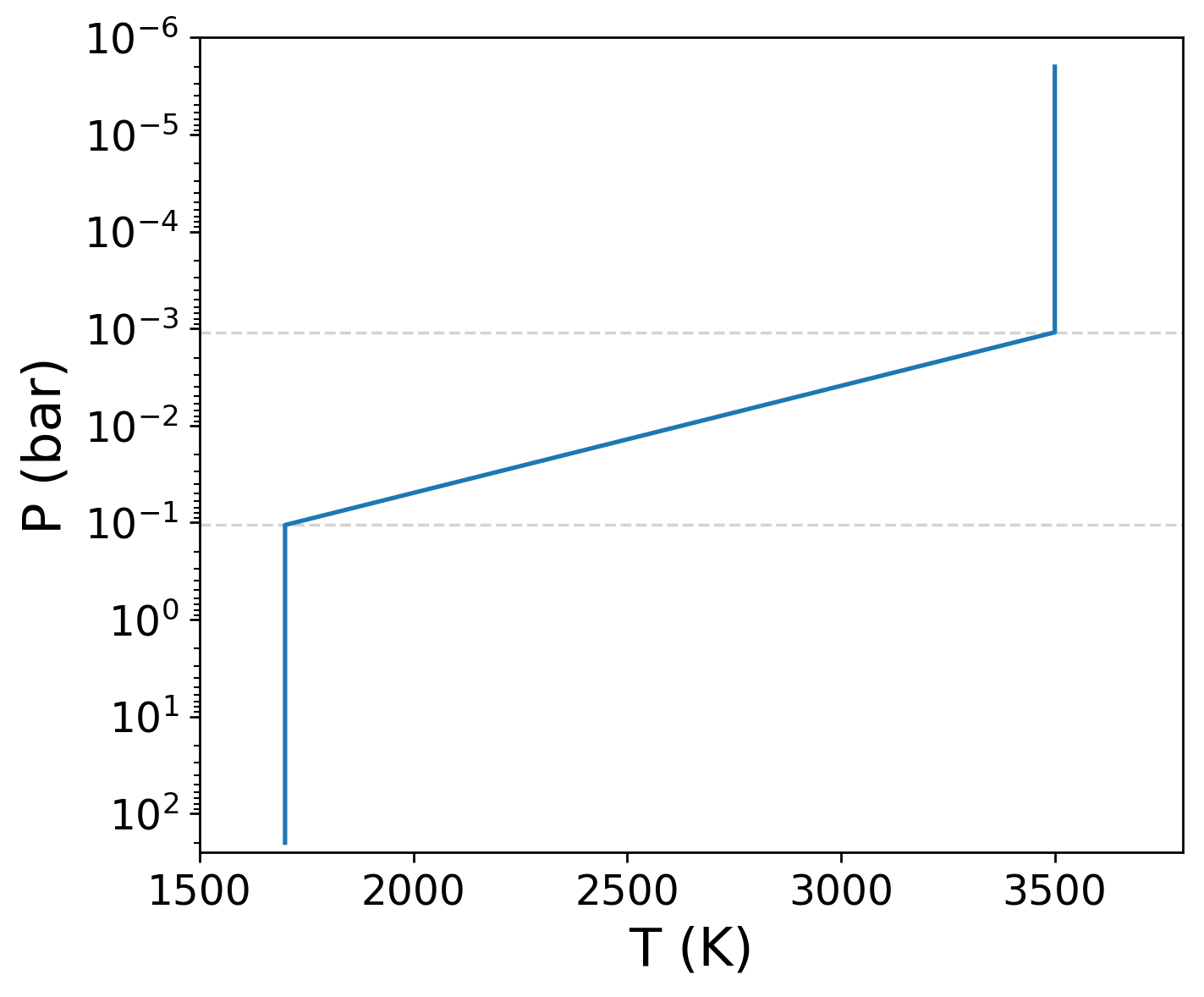}
    \includegraphics[width=0.55\hsize]{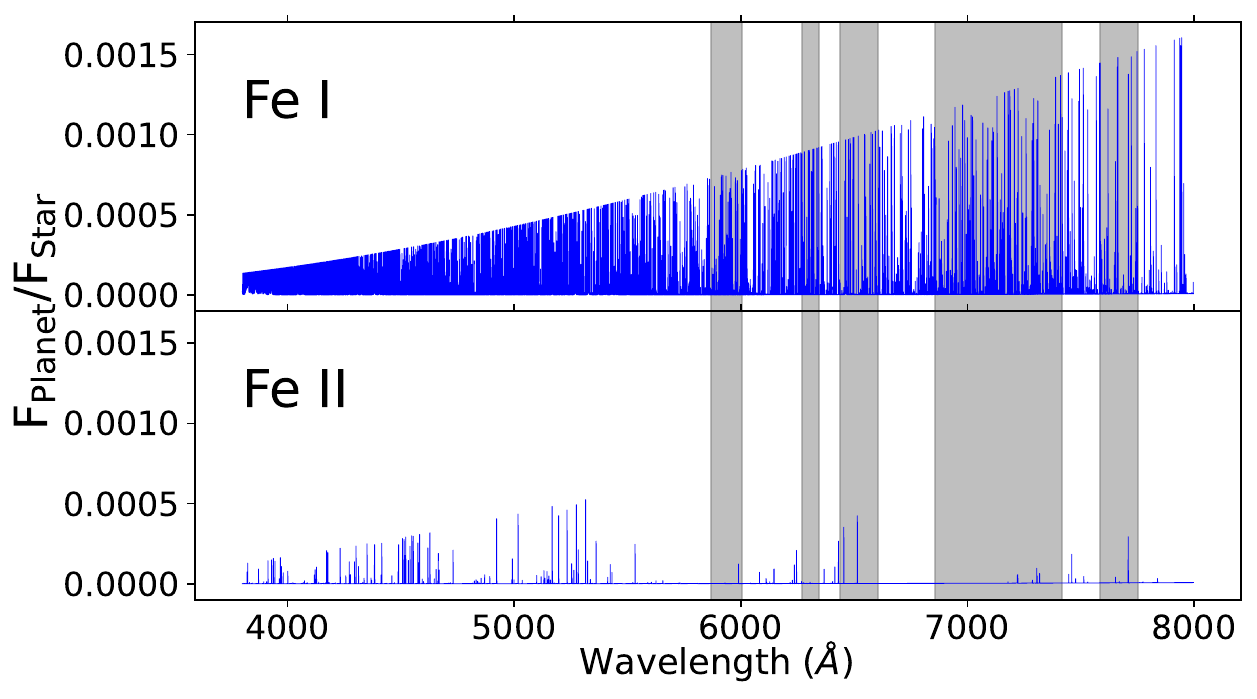}
    \caption{Details of models created for the CCF computation. \textbf{Left:} Pressure-temperature (P-T) profile assumed for the dayside of \planet\ to create the templates of \feI\ and \feII, hereafter PT01 (see Sect. \ref{synthetic_models}), based on the GCM work of \citet{wardenier2021, wardenier2023}. \textbf{Right:} petitRADTRANS models for \feI\ (\textit{top}) and \feII\ (\textit{bottom}), assuming the thermal profile PT01. The shaded regions are excluded from the CCF calculation due to telluric residuals on the empirical spectra.}
    \label{fig:prt_models}
\end{figure*}

    \item Co-add CCFs in the planet rest frame.\\
    The last step is to shift the individual CCFs to the planetary rest frame, sum them, and assess if there is a detection. The planetary signal has a small amplitude in the individual CCFs, but given that the CCF peaks are expected to align when working in the planet's rest frame, then the summation can provide detections at a higher confidence level. We shift the CCFs according to the Keplerian motion of the planet, computed using the parameters shown in Table \ref{tab:properties}. Our analysis is twofold: firstly, we analyse the co-added CCF resulting from each independent epoch (we combine the 39 CCFs of epoch I into one, the 35 CCFs of epoch II into one, and so on); secondly, we construct a co-added CCF without the separation of epochs, so we combine the 148 CCFs into one. A detection of emission lines will manifest itself as a positive peak at (or close to) RV\,$=$\,0\,km/s, since we are analysing it in the planet's rest frame.

\end{itemize}

\subsection{Synthetic models for CCF}
\label{synthetic_models}

The synthetic models to which we compared the observations were produced with the Python package \texttt{petitRADTRANS} (version 2.7.7) \citep{molliere2019, molliere2020, alei2022}, which can calculate both transmission and emission spectra of exoplanets. We chose the high-resolution mode ("lbl", $\lambda/\Delta\lambda=10^6$) to better match our observations, and created separate emission templates with the spectral lines of \feI\ and \feII\ (opacities were contributed to \texttt{petitRADTRANS} by K. Molaverdikhani\footnote{https://petitradtrans.readthedocs.io/en/latest/content/\\available\_opacities.html\#contributed-atom-and-ion-opacities-high-resolution-mode}, calculated from the line lists of R. Kurucz\footnote{http://kurucz.harvard.edu/}). We set the planetary parameters to those of \planet, assuming a hydrogen-helium atmosphere with iron as the only trace species. We set both the \feI\ and \feII\ abundances to be the solar abundance of \feI\ \citep{lodders2020}, keeping this value constant in every layer of the atmosphere. The P-T profile was based on the GCM work of \citet{wardenier2021, wardenier2023} (shown in Fig. \ref{fig:prt_models}). We define a two-point model, where the deep and outer atmospheres are represented by isotherms at 1700\,K and 3500\,K, respectively. The inversion layer is described by a gradient between 1\,$-$\,100 mbar that connects the two isotherms. At a later stage, we diverge from this model and define new atmospheric profiles to evaluate the effect it has on the CCF signature.

The package \texttt{petitRADTRANS} provides spectra in units of spectral flux density (erg cm $^-2$ s$^-1$ Hz$^-1$). We transform them into contrast models (F$_\mathrm{planet}$/F$_\mathrm{star}$) by multiplying by the area of the planet disk and dividing by the flux of the star (modelled as a blackbody with T$_\mathrm{eq} = 6329\,$K. Figure \ref{fig:prt_models} displays the models of \feI\ and \feII\ used in the CCF calculations. The models are interpolated to the same wavelength grid as the observations when computing the CCFs.

\section{Results} \label{results}


\begin{figure*}[h!]
    \centering
    \includegraphics[width=0.49\hsize]{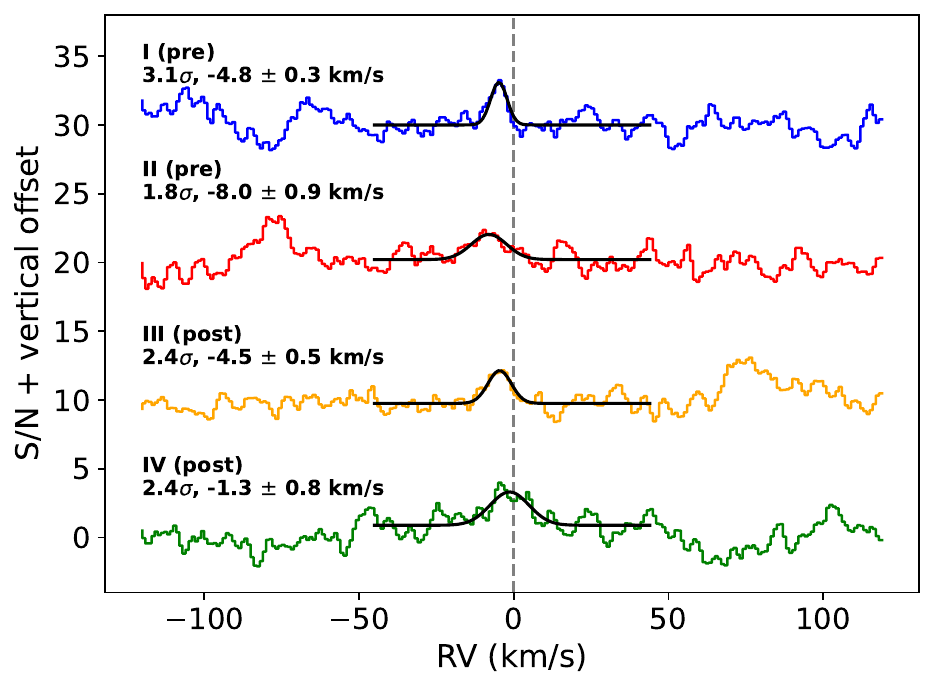}
    \includegraphics[width=0.49\hsize]{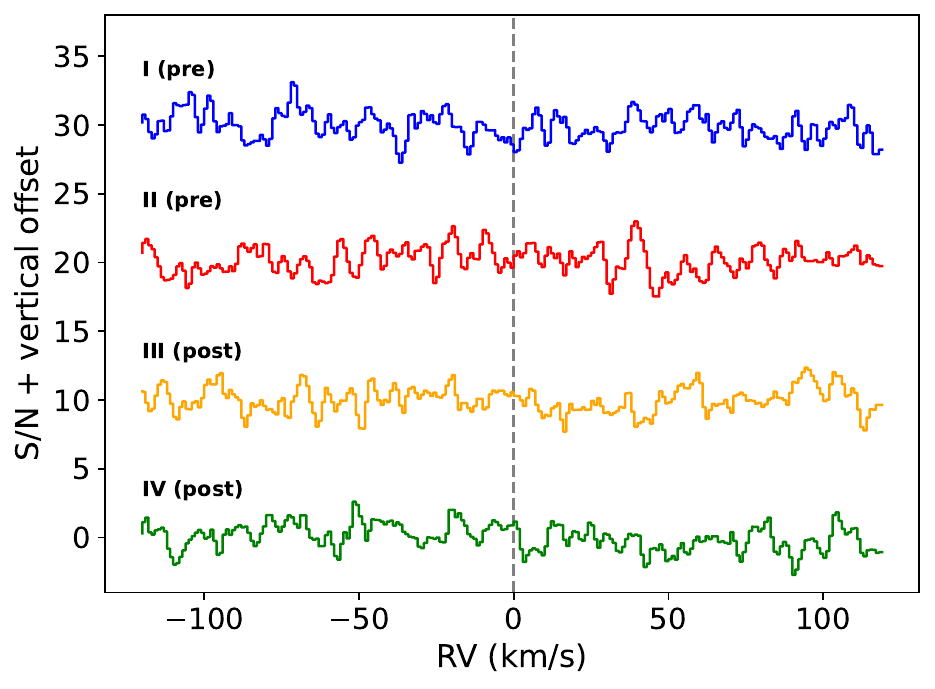}
    \includegraphics[width=0.49\hsize]{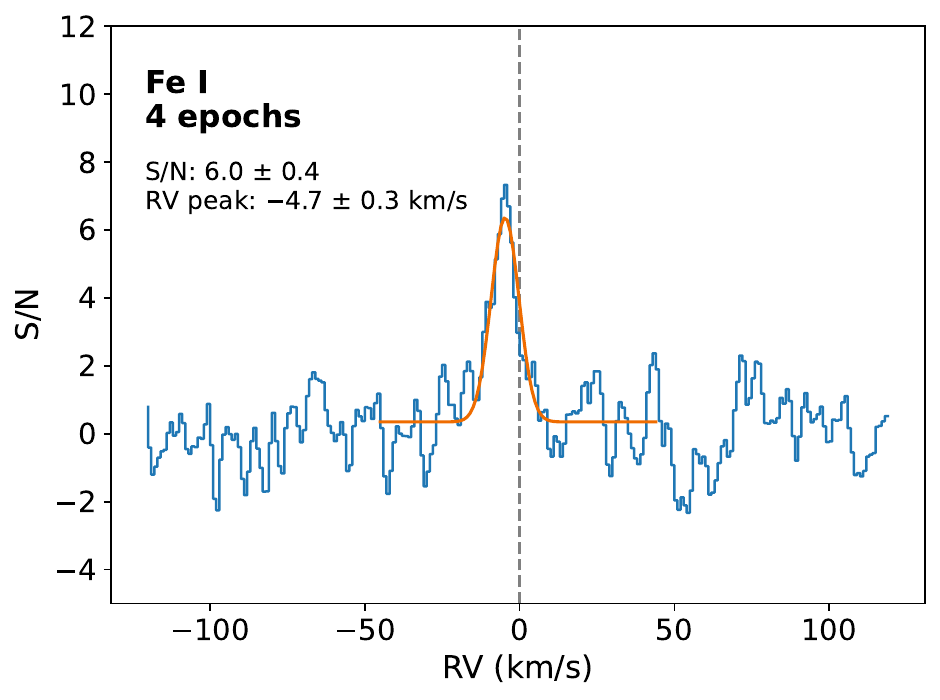}
    \includegraphics[width=0.49\hsize]{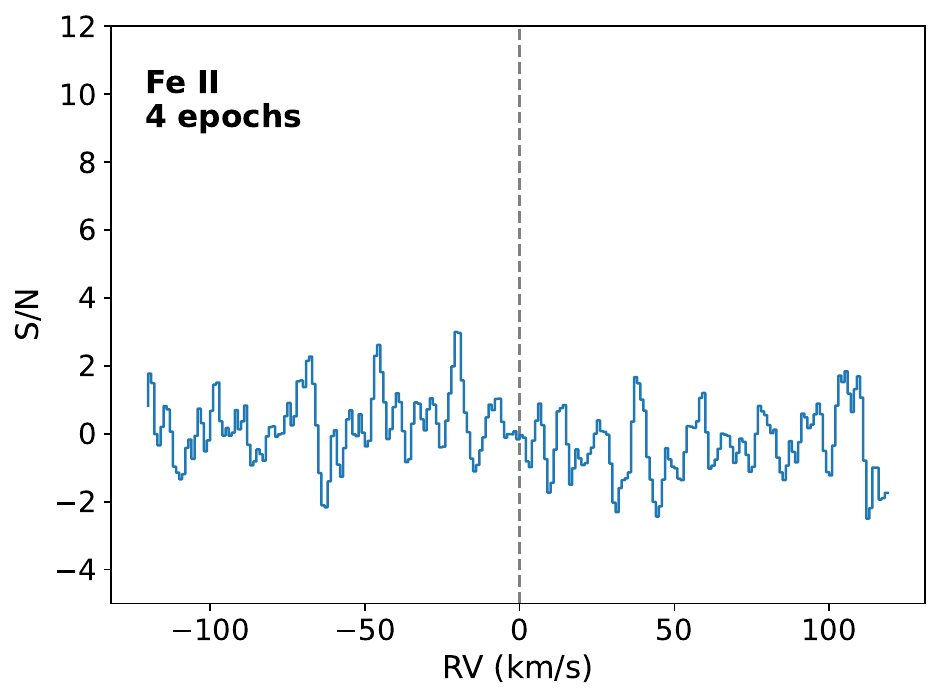}
    \includegraphics[width=0.49\hsize]{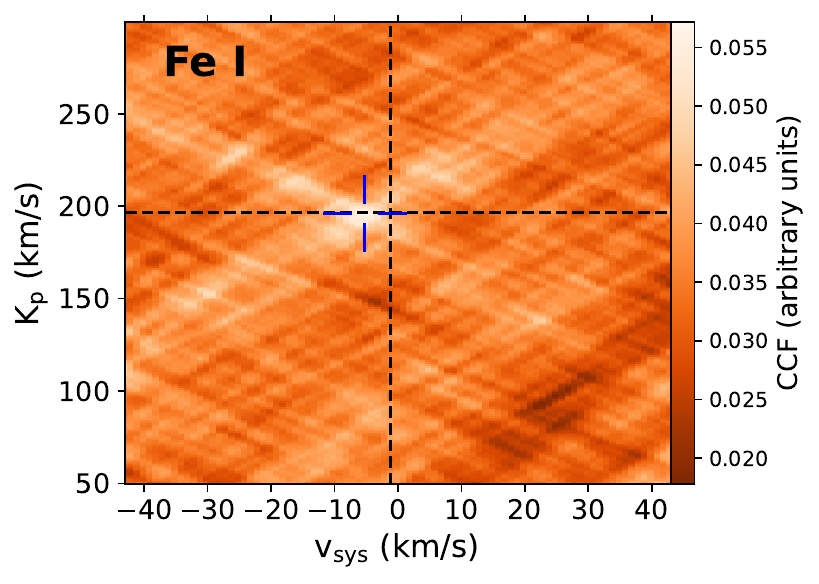}
    \includegraphics[width=0.49\hsize]{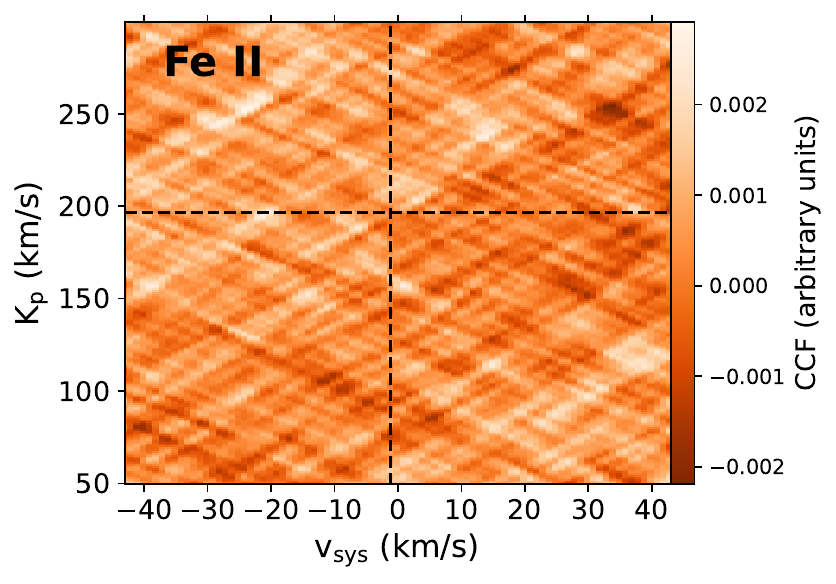}
    \caption{Results of the CCF analysis for \feI\ (\textit{left}, detection) and \feII\ (\textit{right}, non-detection). \textbf{Top:} Summed CCFs in the planet rest frame, separated by epoch, with 1D Gaussian fits (black lines). \feI\ is tentatively detected at a 2$-$3$\sigma$ level in each epoch and all epochs show a blueshifted signal. \textbf{Middle:} Summed CCFs of the four epochs in the planet rest frame, with no epoch separation. \feI\ shows a 6.0$\sigma$ detection, as traced by the 1D Gaussian fit (black line). \textbf{Bottom:} \kpvsys\ plot of the four epochs combined (see Fig. \ref{fig:kp_vsys_epochs} for the individual epochs). The black dashed lines indicate the expected position of the signal. The location of the strongest signal is pinpointed by the blue dashes, for the case of \feI.}
    \label{fig:ccf}
\end{figure*}

The resulting CCFs for neutral and ionised iron are presented in Fig. \ref{fig:ccf}. In the top panels, the CCFs have been co-added in the planet's rest frame for each epoch, calculating the RV shift with the Keplerian solution defined using the parameters presented in Table \ref{tab:properties}. The CCFs were then converted into a \snr\ scale by calculating the standard deviation of the baseline (between [$-$115, $-$35]\,km/s and [35, 115]\,km/s) and dividing everything by this measurement. Furthermore, we fit a simple 1D Gaussian curve locally around zero to retrieve the RV value of the signature. In the middle panels, we show the CCF resulting from adding all curves from the four epochs in the planet's rest frame (normalised by the baseline after the summation), also fitted with a Gaussian curve. 
Lastly, in the bottom panels, we produce the \kpvsys\ plot. For this, we add all 148 CCFs in the planet's rest frame, but for each row of the plot, we compute the Keplerian solution assuming a different value for the semi-amplitude velocity, \kp\ (all other parameters remain constant, see Table \ref{tab:properties}). If the theoretical orbital parameters are correct and no atmospheric dynamics are detectable in the exoplanet, we should expect a detection at the intersection of \kp\,$=\,$196.52\,km/s and \vsys\,$\sim$\,$-$\,1.2\,km/s (traced by the dashed lines in the figure).


\begin{table*}[h]
    \caption{Results from the Gaussian fits to the CCFs.}
    \label{tab:ccf_significance}
    \centering
    \begin{tabular}{m{1cm}>{\centering}m{4cm}>{\centering}m{1.8cm}>{\centering}m{1.5cm}>{\centering}m{3cm}>{\centering\arraybackslash}m{2.5cm}}
         \hline \hline
         Species & Epoch & Detection & \snr\ & RV peak (km/s) & FWHM (km/s) \\
         \midrule 
         \feI         & I                                    & yes          & 3.1$\sigma$ & $-$4.8\,$\pm$\,0.3 & 6.3 \\
                      & II                                   & tentative    & 1.8$\sigma$ & $-$8.0\,$\pm$\,0.9 & 13.5 \\
                      & III                                  & tentative    & 2.4$\sigma$ & $-$4.5\,$\pm$\,0.5 & 8.8  \\
                      & IV                                   & tentative    & 2.4$\sigma$ & $-$1.3\,$\pm$\,0.8 & 14.8 \\
            
                      & I \& II (pre-eclipse)                & yes          & 3.8$\sigma$ & $-$6.0\,$\pm$\,0.4  & 9.4 \\
                      & III \& IV (post-eclipse)             & yes          & 3.5$\sigma$ & $-$3.3\,$\pm$\,0.5  & 11.9 \\
                      & Four epochs co-added                    & yes          & 6.0$\sigma$ & $-$4.7\,$\pm$\,0.3  & 10.6 \\
                      \midrule 
         \feII & I, II, III, IV    & no  & - & - & - \\
         \hline
    \end{tabular}
\end{table*}

\subsection{Detection of blueshifted Fe\,I in emission}
\label{iron_detection}

We report a 3.1\,$\sigma$ detection of \feI\ in epoch I, and tentative detections in epochs II, III, and IV, with significances of 1.8\,$\sigma$, 2.4\,$\sigma$, and 2.4\,$\sigma$, respectively. The \snr\ of these detections is rather low, and other peaks are visible in the co-added CCFs. 
Some of these peaks lie slightly above what could be considered the noise level, at values distant from the zero-velocity point. The CCF process itself can introduce artefacts resembling detection peaks when lines from the template randomly match with the empirical features of other species scattered throughout the planet's spectra \citep{borsato2023}. However, in this case, it seems that the noise in the continuum is mostly dominated by red noise.

The weakest detection of the individual epochs comes from epoch II, which is puzzling at first because these observations had the best seeing conditions of the four epochs observed, and the spectra have the best \snr. However, this was the only epoch observed with UT2 of the VLT, whereas UT1 was used for the other epochs. UT2 tends to be more affected by the interference pattern created in ESPRESSO (wiggles), therefore the correction of this effect might have left more residuals compared to the other epochs, in turn leading to a less significant detection. \citet{prinoth2023} reported similar quality issues between data from UT1 and UT2 (see their Appendix A).

When combining all epochs, the neutral iron detection is much clearer, at a \snr\ of 6.0\,$\sigma$ (see Fig. \ref{fig:ccf}). Table \ref{tab:ccf_significance} summarises the significance of our findings. Lastly, in the \kpvsys\ plot (Fig. \ref{fig:kp_vsys_epochs}), \feI\ is detected at the expected \kp\ value, with a blueshifted \vsys, and no other strong peaks are found in the explored parameter space.
The detection of emission lines confirms the existence of a thermal inversion layer in the dayside of \planet. The inversion had been previously hinted at by \citet{edwards2020, may2021, fu2021}, and confirmed by \citet{yan2023} using near-infrared CRIRES+ emission spectroscopy. Our data support their findings.

Furthermore, we report that our \feI\ detection is blueshifted with respect to the planetary rest frame. The CCF for the four epochs produces a peak at an RV of $-$4.7\,$\pm$\,0.3\,km/s (full width at half maximum, FWHM, of 10.6\,km/s). For the co-added CCFs of each epoch, the Gaussian fits are centred at $-$4.8\,$\pm$\,0.3, $-$8.0\,$\pm$\,0.9, $-$4.5\,$\pm$\,0.5, and $-$1.3\,$\pm$\,0.8\,km/s, with FWHM values of 6.3, 13.5, 8.8, and 14.8\,km/s, respectively for epochs I, II, III, and IV (see Fig. \ref{fig:ccf}). We note that there is some scatter in the RV shift observed from epoch to epoch, though it is unclear if the differences are caused by the low \snr\ of the planetary signature or if they indicate a physical variation of the atmosphere. 
Moreover, our uncertainties are likely underestimated as they represent only the nominal error of the Gaussian fit. It is difficult to properly account for the uncertainties created by the correlated noise of the CCF. Notwithstanding, all the observations presented have clear blueshifted peaks that are identified consistently across the epochs.
There seems to be a drift of $\sim$3\,km/s between the observations of 2021 (II and III) and 2022 (I and IV) for the same phases. However, it is not possible to draw any significant conclusions at this point. 
In Fig. \ref{fig:ccf_pre_post_eclipse}, we present the combined CCFs by phase range, for pre- or post-eclipse. The observations after the eclipse reveal a smaller shift, $-$3.3\,$\pm$\,0.5\,km/s (FWHM\,$=$\,11.9\,km/s) than those obtained before the eclipse, $-$6.0\,$\pm$\,0.4\,km/s (FWHM\,$=$\,9.4\,km/s). The observations of epochs I and II (pre-eclipse) only overlap partially in phase coverage, so this could indicate a gradual change as the planet rotates. However, even with epochs covering very similar phase ranges, such as epochs III and IV, the RV peaks of their CCFs are discrepant at the 5$\sigma$ level. The Gaussian fits on epochs I (pre) and III (post) show less broadening than for II (pre) and IV (post), so we consider the fits of I and III to be more significant than their counterparts at the same phase. This implies a consistent blueshift in the emission signal from both the east and west dayside hemispheres of \planet\ of about $-$4.6\,km/s (see Fig. \ref{fig:ccf}). Overall, further observations would be useful to confirm if these velocity discrepancies are the result of physical processes, and to shed more light on what could be causing them.
 
We considered if adopting an eccentric orbit, instead of a fully circular orbit, would alter the results and eliminate the blueshift. Given the age of the system \citep[1.8\,Gyr,][]{ehrenreich2020}, we expect the orbit to have circularised (see favourable arguments for this orbital solution in \citeauthor{ehrenreich2020} \citeyear{ehrenreich2020}, Methods, but see also \citeauthor{valentecorreia2022} \citeyear{valentecorreia2022}). Thus we assumed e$\,=\,$0 in the calculations described so far. However, a slightly eccentric orbit might perhaps explain this signature. Furthermore, \citet{savel2022} reported that allowing for a small eccentricity of 0.01 was a necessary adjustment to reproduce the transit signature of \feI\ in \planet\ \citep{ehrenreich2020}, combined with high-altitude, optically thick clouds of \feI, Al$_2$O$_3$, and Mg$_2$SiO$_4$. So we investigated the possibility of \planet\ having an eccentric orbit.

\begin{figure}
    \centering
    \includegraphics[width=\hsize]{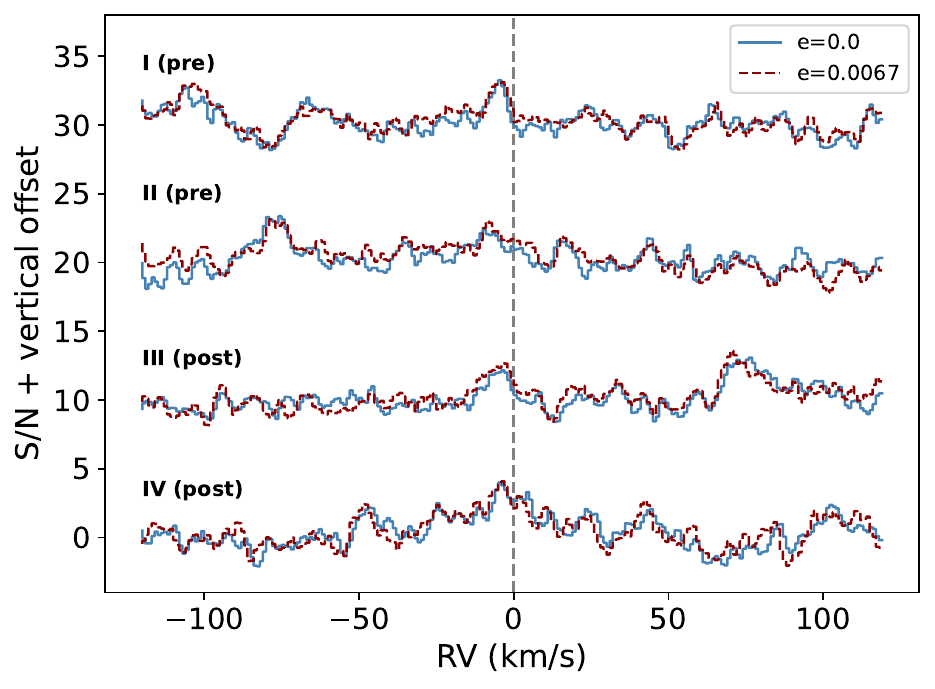}
    \includegraphics[width=\hsize]{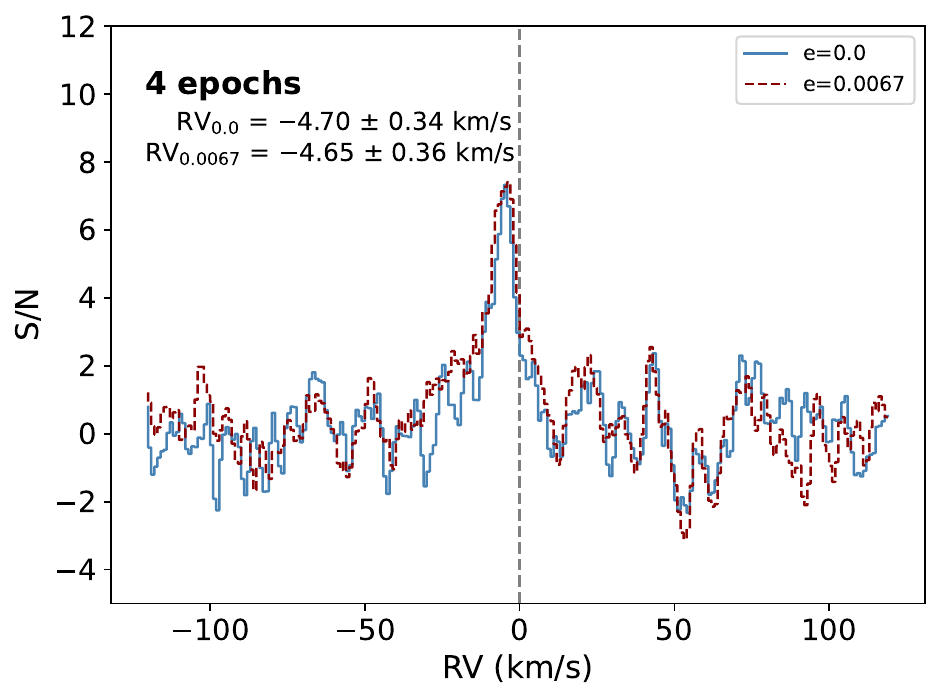}
    \caption{CCFs of \feI\ calculated for the cases of e=0 and e=0.0067 (CHEOPS upper limit, \citeauthor{demangeon2024} \citeyear{demangeon2024}). The CCFs in the top panel are separated by epoch, and the differences in the RV peaks between these eccentricity scenarios are between 0 and 3\,km/s. In the bottom panel, all epochs are co-added, and the difference between the two peaks is negligible. Assuming a slightly eccentric orbit, rather than a circular orbit, does not eliminate the blueshifted signature seen for \feI.}
    \label{fig:ecc_effect}
\end{figure}

The phase curve of \planet\ was recently observed with CHEOPS\footnote{CHaracterising ExOPlanets Satellite} \citep{demangeon2024}. These authors placed an upper limit on the eccentricity of e$\,=\,$0.0067. We set the eccentricity to this upper limit, computed the CCFs with the newly shifted spectra, and compared the two cases. In Fig. \ref{fig:ecc_effect}, we present the co-added CCFs for the cases of e$\,=\,$0 and e$\,=\,$0.0067. The change in RV of the CCF peaks of individual epochs is between 0\,km/s and 3\,km/s (top panel), with the peak in epochs I and II being less blueshifted compared to the zero-eccentricity case, and epochs III and IV being more blueshifted. When all CCFs are co-added (bottom panel), the change is negligible. As the blueshift remains present in all cases, we rule out the possibility that this signature is due to unaccounted-for eccentricity. In Sect. \ref{discussion}, we discuss further possible origins for the blueshifted \feI.

\subsection{Non-detection of Fe\,II in emission}

We did not detect the presence of \feII, even when adding the 148 CCFs from all epochs (see right panels of Fig. \ref{fig:ccf}). \feII\ has been detected in several UHJs via transmission spectroscopy, thus it is expected to be present on the daysides of these planets. However, it is expected to be more abundant at high altitudes, where \feI\ is ionised by the hotter temperatures. These atmospheric layers are more challenging to probe with emission spectroscopy due to being more optically thin. Thus it is not surprising we could not detect \feII\ emission on \planet, even if we expect it to be present.

\subsection{Constraining P-T profile}
\label{sect:pt_exploration}

\begin{figure*}[h]
    \centering
    \includegraphics[width=0.90\hsize]{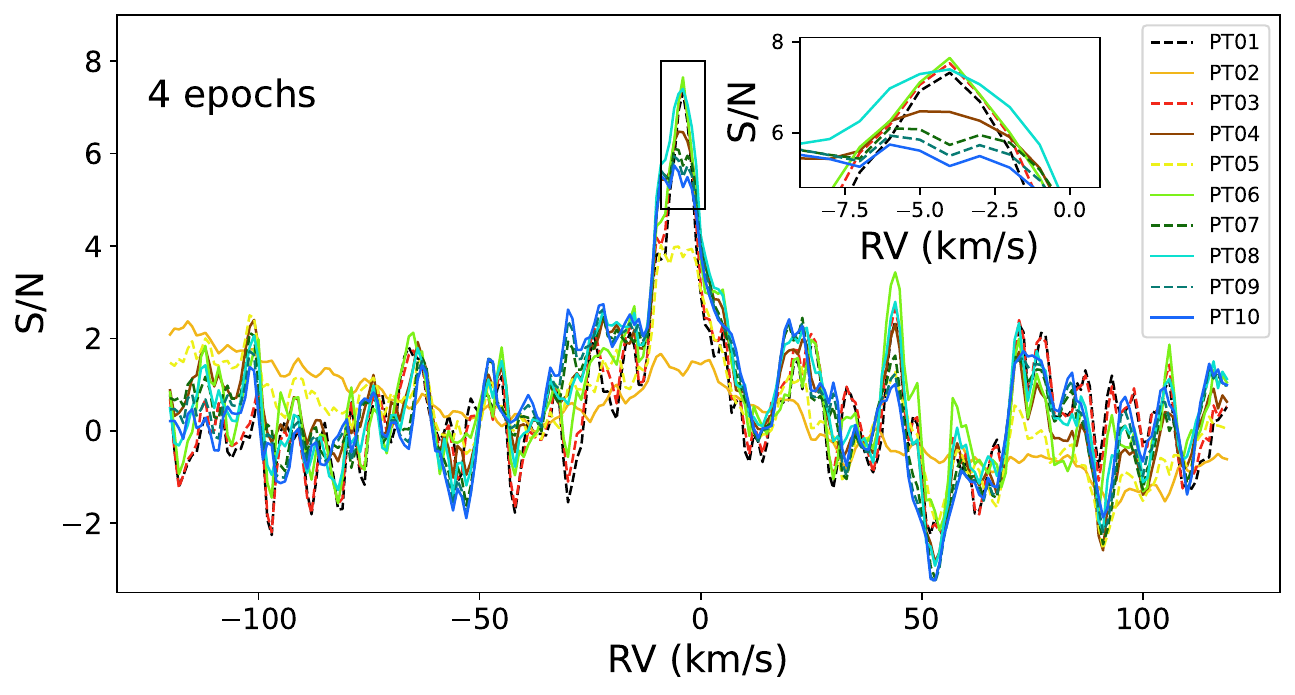}
    \includegraphics[width=0.51\hsize]{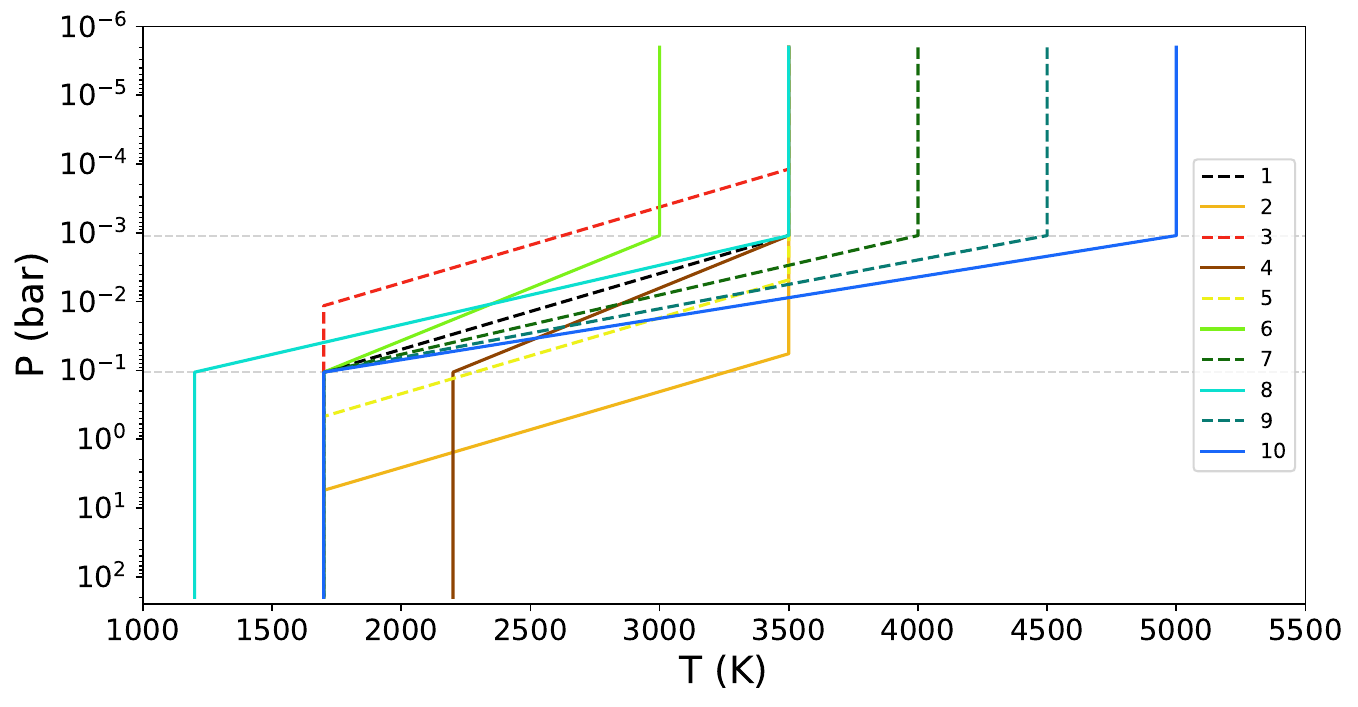}
    \includegraphics[width=0.37\hsize]{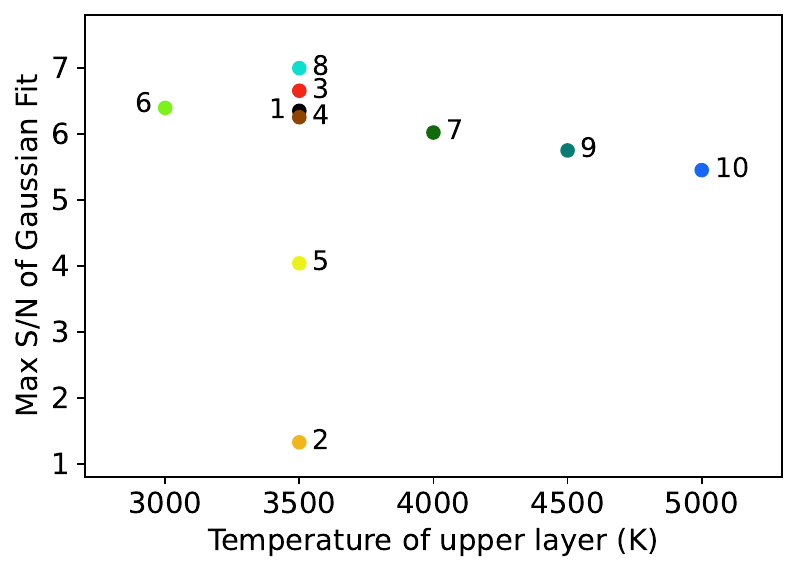}
    \caption{Analysis of the dependence of the iron detection on the model used to compute the CCF. \textbf{Top:} Four-epoch co-added CCFs for \feI\ computed with different templates that correspond to each P-T profile (bottom left panel). \textbf{Bottom left:} P-T profiles (see also Fig. \ref{fig:prt_models_pt_exploration} and Table \ref{tab:tp_details}). \textbf{Bottom right:} Maximum \snr\ value of Gaussian fit to co-added CCFs of four epochs, computed with the \feI\ template corresponding to each P-T profile.}
    \label{fig:tp_exploration}
\end{figure*}

Once we had confirmed the presence of \feI\ in emission, we investigated how changes in the P-T profiles would affect the CCF peak. Thus we defined ten different P-T profiles, computed the synthetic template with \texttt{petitRADTRANS}, and calculated the cross-correlation for all epochs. Starting from the initial P-T profile \citep[PT01, see Fig. \ref{fig:prt_models} and Sect. \ref{synthetic_models},][]{wardenier2021, wardenier2023}, we modified either the temperature of the lower layer of the atmosphere, the temperature of the upper layer, the pressure boundaries of the inversion layer, or a combination of these. The characteristics of each P-T profile are shown in the bottom left panel of Fig. \ref{fig:tp_exploration} and listed in Table \ref{tab:tp_details}. The corresponding synthetic templates that were utilised for the CCF can be found in Fig. \ref{fig:prt_models_pt_exploration}. Despite the fact the \feI\ opacities in \texttt{petitRADTRANS} were only calculated for temperature values up to 4000\,K, we decided to go above this value in some of the P-T profiles. This means that the opacities used by \texttt{petitRADTRANS} when T$>$4000\,K are the same ones as for T$=$4000\,K. However, the radiative source function and the atmospheric scale height are not constant for temperature values that extend beyond the pre-defined P-T grid, thus they still hold some useful information.

We co-added the 148 CCFs corresponding to each P-T template (top panel of Fig. \ref{fig:tp_exploration}) and fitted the resulting peaks with Gaussian curves. The \feI\ detection can be found in all tested profiles except for PT02, which corresponds to the profile with the inversion temperature located deep in the atmosphere. As expected, the significance of the detection changes in the different cases, with the weakest detection stemming from PT05, another profile that assumes the inversion temperature deeper in the atmosphere, though not as deep as for P0T2.  

Modifying the temperature of the lower or upper layers seems to have a limited impact on the significance, as the values only increase or decrease slightly compared to the base model PT01. The case that produces the strongest detections is the one where the temperature of the lower layer was reduced to T$_\mathrm{inner}\,=\,1200$\,K.

A further visual investigation of the models revealed that many lines were saturated. This is likely due to the cut-off temperature defined for the isotherm at lower pressures. Strong lines that reach the blackbody curve defined by this isotherm thus have the same brightness temperature and appear saturated. In principle, if the lines were not saturated, the significance of the \feI\ detection would change, but it would not alter our ultimate scientific conclusion that neutral iron is detected on \planet.

\begin{figure}
    \centering\includegraphics[width=\hsize]{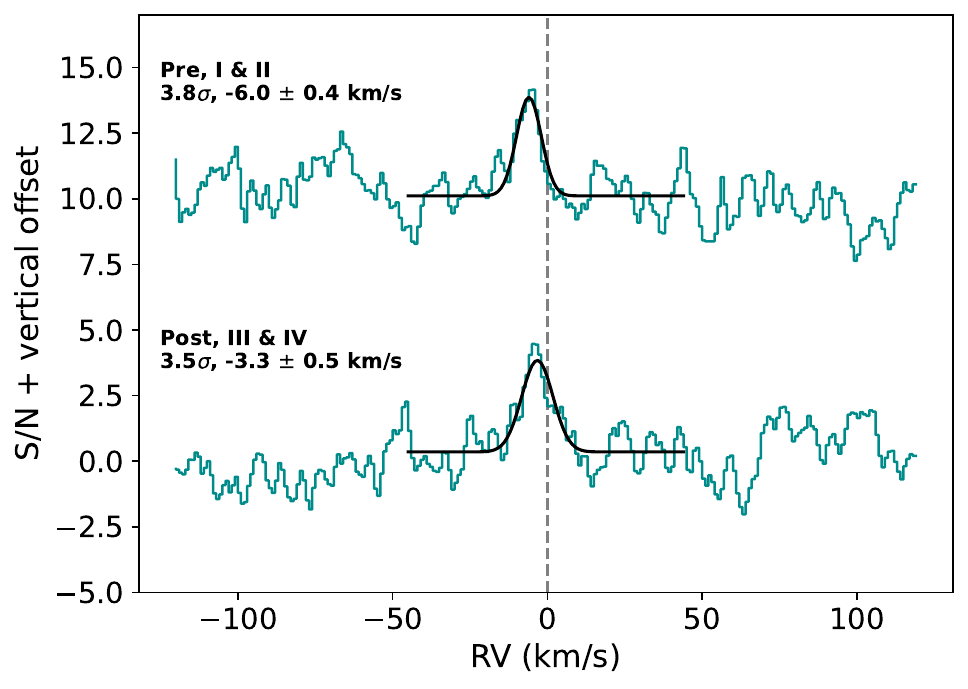}
    \caption{Summed CCF curves of the epochs before the eclipse (\textit{top}) and after (\textit{bottom}). The \feI\ signal is more blueshifted in phases before the eclipse, and becomes less blueshifted after.}
    \label{fig:ccf_pre_post_eclipse}
\end{figure}


\section{Discussion} \label{discussion}

\begin{figure*}[h]
    \centering
    \includegraphics[width=0.9\hsize]{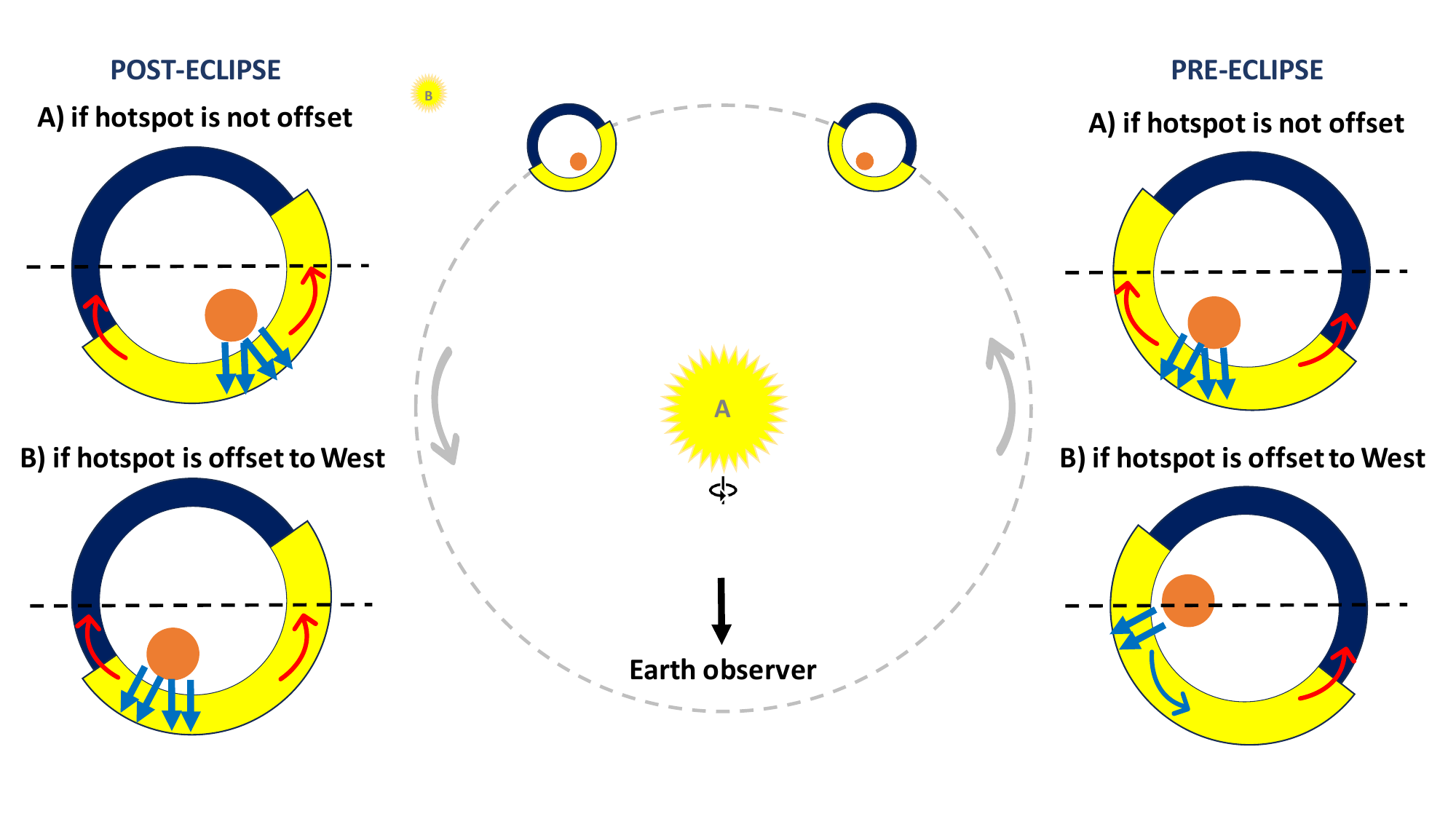}
    \caption{Geometry of the \planet\ system. The planet is shown from a polar perspective, at different phases of its orbit, with the inflated dayside in yellow and the nightside in dark blue. The orange circle is the hotspot, for which we consider two possible locations: (A) at the substellar point (no offset) or (B) with a westward offset; to compare the possible atmospheric dynamics of each case. The arrows represent the motion of winds, and whether they appear blueshifted or redshifted to an observer on Earth. The dashed black line separates the hemisphere visible to the observer from the non-visible one. A proposed scenario of atmospheric circulation that could lead to the results of this work is presented in the text (see Sect. \ref{fe_blushift_discussion} for discussion).}
    \label{fig:geometry}
\end{figure*}

\subsection{Blueshifted Fe\,I emission}
\label{fe_blushift_discussion}

The \feI\ emission signature we report on the dayside of \planet\ is blueshifted by $\sim-$4.7\,km/s. Similar blueshifted signals have been identified in recent works, such as in CO and H$_2$O in WASP-77A\,b \citep{line2021}, and H$_2$O in our target, \planet\ \citep{yan2023}. However, the mechanism behind them has not been investigated so far. 


The Doppler shift observed here cannot be explained by an eccentricity of the orbit, which is very close to zero (see Sect. \ref{iron_detection}). It also does not trace solely the day-to-night wind proposed in recent transmission studies \citep{seidel2019, seidel2021, ehrenreich2020}, which would appear redshifted in dayside observations (in the planet rest frame). Moreover, this wind is seen at the atmospheric limbs with transit spectroscopy, whereas our observations are most sensitive to integrated dayside emission. It seems to require that additional components be added to the atmospheric dynamics scenario, to ensure that the final disk-integrated signature is blueshifted. In this section, we propose a simple scenario of atmospheric circulation that could explain our observations, and we illustrate the proposed dynamics of the planet's atmosphere in Fig. \ref{fig:geometry}.

Our interpretation is that material on the dayside of \planet\ is moving towards the observer, with similar magnitude on both the east and west hemispheres. The day-to-night heat redistribution pattern proposed by GCMs for UHJ in the presence of drag \citep[e.g.][]{wardenier2021} would have an RV component close to zero at orbital phases close to the eclipse, when the dayside disk is almost perpendicular to the observer's perspective. 

In emission spectroscopy, observations are more sensitive to the hottest region of the dayside, in the vicinity of the substellar point. As such, we could be detecting \feI\ atoms that are rising in the atmosphere, in a radial motion from the inner to the outer atmospheric layers. This displacement is possibly generated by the hotspot at or close to the substellar point. A series of works by \citet{seidel2019, seidel2020, seidel2021} has shown that vertical upwards winds in the upper atmosphere of UHJs, including \planet, were the likely cause for the broadening observed in the Na\,I doublet, being of the order of 20\,km/s. Moreover, if these hotter parcels of atmosphere are transported from the dayside to the nightside of the planet, via the day-to-night wind or the super-rotating equatorial jet, the atoms would be detectable at the terminator via transmission spectroscopy, as was reported by \citet{ehrenreich2020, kesseli2021, kesseli2022, pelletier2023}. Furthermore, if these atoms condense when reaching the nightside, due to the cooler temperatures, it could explain the glory effect reported by \cite{demangeon2024} from the light curve analysis.

This could mean that the hotspot is generating the upward displacement of a substantial amount of \feI\ atoms. \citet{sainsbury-martinez2023} investigated GCMs models that included vertical transfer of heated material in the outer atmosphere of UHJs. They concluded that the differences between the day and nightside temperatures ultimately lead to transport in the upward direction on the dayside, whereas the nightside sees a downward motion. However, these authors assumed a P-T profile that combined an adiabat for the deep atmosphere and an isotherm for the outer layers, with no inversion layer present. Had a thermal inversion been included, we do not know to what extent it would alter their findings. 

As mentioned in the previous section, there is some doubt as to whether the changes in RV shift from one epoch to another are real, or if they are a product of the small \snr\, residual contamination from activity, or a combination of these, for example. How (or if) the magnitude of the shift varies can help us locate the hotspot of \planet. If we consider that the planetary signal is more blueshifted before the eclipse than after (see Fig. \ref{fig:ccf_pre_post_eclipse}), that could mean the hotspot is shifted to the west. As exemplified in Fig. \ref{fig:geometry}, a westward shift would create a stronger blueshift signature before the occultation. Whereas if we analyse the CCFs for each epoch separately, one can argue that the peaks on epochs II and IV have a less constrained fit than those of epochs I and III, thus we should only consider the two latter results in our discussion. In this case, the blueshift has the same magnitude across both phases, which can point to the lack of an offset for the hotspot. With a hotspot that is located at or close to the substellar point, observations before and after the occultation would reveal a mirrored wind structure and thus create a similar atmospheric RV shift overall. \citet{may2021} have reported a negligible offset of the hotspot for \planet, though this conclusion was drawn from Spitzer data which is potentially probing a different altitude in the atmosphere. \citet{beltz2022a} show that the hotspot offset can be reduced as a result of applying more sophisticated active magnetic drag treatments over more approximate ones. This goes to show the importance of studying planetary magnetic fields and their impact on atmospheric circulation. On the other hand, \citet{wardenier2021} and \citet{savel2022} required a hotspot offset in their GCMs in order to reproduce the observational findings of transmission spectroscopy. More observations of the dayside of \planet\, with higher \snr\ and better time resolution are necessary to constrain the position of the hotspot.

To further dive into the atmosphere of \planet, it would be necessary to develop a retrieval framework, such as those presented in \citet{brogi2019, seidel2020, seidel2021, pelletier2021, gandhi2022}, adapted to emission spectroscopy. Such a task is outside the scope of this paper. GCM studies are also an efficient tool to unravel the underlying atmospheric phenomena at play in the atmospheres of UHJs. \citet{wardenier2021, wardenier2023} have developed a 3D model for \planet, but have produced only the transmission spectra for comparison with the already published transit data \citep{ehrenreich2020}. Producing GCM to delve into the scorching dayside of this planet is, likewise, not the goal of this observational paper, though we strongly encourage this effort.

\subsection{Lack of Fe\,II detection}
\label{sect:feII_discussion}


\feII\ has only been detected in the dayside of one other UHJ, KELT-20\,b/MASCARA-2\,b \citep{borsa2022}, and only in the post-eclipse data. Despite this, follow-up studies were not able to find its signature on the same planet \citep{yan2022, kasper2023, petz2024}. A non-detection has also been reported for the dayside of KELT-9\,b by \citet{pino2020} and \citet{riddenharper2023}. \citet{cont2022} did not detect \feII\ on the emission spectra of WASP-33\,b, and their injection-recovery tests concluded that it would not be detectable in their data. For the case of WASP-121\,b, a similar planet to \planet, it was observed with ESPRESSO in eight different epochs, producing about two times as much spectra than we analysed in this work \citep{hoeijmakers2024}, but \feII\ remained undetected in the dayside of this UHJ.

\feI\ ionises at lower pressures and higher temperatures in the atmosphere, decreasing its abundance and increasing that of \feII. However, the upper layers of the atmosphere are more optically thin. This is the reason why transmission spectroscopy is a better method to explore them and has been a more successful technique at finding \feII. To probe \feII\ at these pressures with emission spectroscopy, it would require a greater abundance of this ion, or a larger number of strong lines available to probe it. Our \feII\ template in the optical regime contains $\sim$1200 lines, which is one order of magnitude smaller than the $\sim$11700 lines present in the \feI\ template, a species that is more abundant further down in the atmosphere, and thus more amenable to be detected in thermal emission. The \feII\ lines become stronger for UHJs with higher equilibrium temperatures, which might make this ion traceable. The aforementioned works investigated planets hotter than \planet, and did not report detections. With \planet\ sitting on the colder edge of the UHJ temperature range, it is expected that we cannot prove the existence of \feII\ on its dayside with CCF techniques.

The lack of confident detections or non-detections of \feII\ does not allow for any meaningful conclusions regarding a population trend. We stress that it would be of great interest to trace the ionised state of iron as we expect it to be affected by planetary magnetic fields.

\section{Summary and conclusions} \label{summary}

We observed the dayside of the ultra hot Jupiter \planet\ with ESPRESSO on four different epochs. We collected a total of 148 high-resolution emission spectra. Half of these were obtained just before the planet's secondary transit (phases 0.34\,-\,0.47), and the other half right after (0.54\,-\,0.62), providing insight into both the east and west hemispheres. This is the first emission spectroscopy study carried out for \planet\ at visible wavelengths. We also present monitoring data of \hoststar\ from EulerCam, which shows that the host star (and its binary companion) are both quiet stars, with little photometric variation.

Our main goal was to detect \feI\ and \feII\ in emission on the dayside of \planet. We used the CCF method to compare the observational data with synthetic models of these chemical species computed with \texttt{petitRADTRANS}. Furthermore, detecting emission features confirms the existence of an inverted atmospheric profile. We then investigated how the pressure-temperature profile impacted the emission signature, by computing synthetic templates based on varying P-T profiles and comparing the resulting CCFs. Our results are summarised as follows:

\begin{enumerate}
    \item We detect a blueshifted signature (-4.7$\,\pm\,$0.3\,km/s) of \feI\ in nearly all epochs, with a detection significance of 6.0\,$\sigma$ from the co-added CCF of the four epochs.
    \item We confirm the existence of a thermal inversion layer, which follows from the fact that emission features can only be present if the dayside has an inverted structure.
    \item We report a non-detection of \feII. Due to the hot temperature of this planet, we expect this ion to exist in the outer atmosphere. However, the non-detection could be due \feII\ being more abundant in the outer atmosphere, which is optically thinner and thus harder to probe with emission spectroscopy. Follow-up studies are required to confirm it its presence.
    \item We discuss possible atmospheric scenarios to explain the blueshifted signature. We propose that material is being radially ejected from the hotspot, rising in the atmosphere, and proceeding to the cooler nightside of the planet. We strongly encourage the development of GCM studies that could reproduce this feature.
    \item Based on the change of RV shift measured in individual epochs, we are unable to constrain whether the hotspot is located at the substellar point or if it is offset. Further observations are required to disentangle the two scenarios.
    \item The \feI\ signal strength changes when we compare the CCFs resulting from different assumptions of P-T profile. The strongest significance is attributed to a profile where the temperature of the upper atmosphere is 3500\,K and the lower atmosphere is 1200\,K, with the inversion located between 1$-$100\,mbar.
\end{enumerate}

In recent years, emission spectroscopy has often been seen as ill-favoured compared to transmission spectroscopy due to the glaring difference in signal strength. However, it is the best spectroscopic avenue to probe the atmospheric structure and composition of exoplanets' dayside, thus being a crucial technique for exoplanet research. In the lead-up to the next generation of ground-based spectrographs, such as ANDES at the Extremely Large Telescope \citep{marconi2022}, we provide an example of how useful ground-based high-resolution instruments are for characterising exoplanets. We highlight, in particular, the capability of ESPRESSO to probe the dayside of ultra hot Jupiters, in order to constrain their chemical composition and probe the dynamics of atmospheric circulation.

\begin{acknowledgements}
We thank the anonymous referee for their comments which helped improve the manuscript. The authors acknowledge the ESPRESSO project team for its effort and dedication in building the ESPRESSO instrument. This work was supported by the Funda\c{c}\~ao para a Ci\^encia e Tecnologia (FCT) and POCH/FSE through the research grants UIDB/04434/2020 and UIDP/04434/2020, and in the framework of the project 2022.04048.PTDC (Phi in the Sky, DOI 10.54499/2022.04048.PTDC). This work was co-funded by the European Union (ERC, FIERCE, 101052347). This project has received funding from the European Research Council (ERC) under the European Union's Horizon 2020 research and innovation programme (project {\sc Spice Dune}, grant agreement No 947634). Views and opinions expressed are however those of the authors only and do not necessarily reflect those of the European Union or the European Research Council. Neither the European Union nor the granting authority can be held responsible for them. The authors acknowledge the financial support of the Swiss National Science Foundation (SNSF). This work has also been carried out within the framework of the National Centre of Competence in Research (NCCR) PlanetS supported by the SNSF under the grants 51NF40\_182901 and 51NF40\_205606. A.R.C.S. acknowledges support from FCT through the fellowship 2021.07856.BD. O.D.S.D. is supported in the form of a work contract (DL 57/2016/CP1364/CT0004) funded by national funds through FCT. C.L. and F.P. would like to acknowledge the SNSF for supporting research with ESPRESSO through the SNSF grants nr. 140649, 152721, 166227, 184618 and 215190. The ESPRESSO Instrument Project was partially funded through SNSF’s FLARE Programme for large infrastructures. H.C. and M.L. acknowledge the support of the SNCF under grant number PCEFP2\_194576. M.R.Z.O. and E.H-C. acknowledge financial support from the Agencia Estatal de Investigaci\'on (AEI/10.13039/501100011033) of the Ministerio de Ciencia e Innovaci\'on and the ERDF "A way of making Europe" through project PID2022-137241NB-C42. E.H-C. acknowledges support from grant PRE2020-094770 under project PID2019-109522GB-C51 funded by the Spanish Ministry of Science and Innovation/State Agency of Research, MCIN/AEI/10.13039/501100011033, and by ERDF, "A way of making Europe". A.S.M. acknowledges financial support from the Government of the Canary Islands project ProID2020010129. A.S.M., J.I.G.H., and R.R. acknowledge financial support from the Spanish Ministry of Science and Innovation (MICINN) project PID2020-117493GB-I00. S.G.S. acknowledges the support from FCT through Investigador FCT contract nr. CEECIND/00826/2018 and POPH/FSE (EC) (DOI: 10.54499/CEECIND/00826/2018/CP1548/CT0002)). S.C. acknowledges financial support from the Italian Ministry of Education, University, and Research with PRIN 201278X4FL and the "Progetti Premiali" funding scheme. C.J.A.P.M. acknowledges FCT and POCH/FSE (EC) support through Investigador FCT Contract 2021.01214.CEECIND/CP1658/CT0001. R.A. is a Trottier Postdoctoral Fellow and acknowledges support from the Trottier Family Foundation. This work was supported in part through a grant from the Fonds de Recherche du Qu\'ebec - Nature et Technologies (FRQNT). This work was funded by the Institut Trottier de Recherche sur les Exoplan\`etes (iREx). T.A.S. acknowledges support from FCT through the fellowship PD/BD/150416/2019. E.P. acknowledges financial support from the Agencia Estatal de Investigaci\'on of the Ministerio de Ciencia e Innovaci\'on MCIN/AEI/10.13039/501100011033 and the ERDF “A way of making Europe” through project PID2021-125627OB-C32, and from the Centre of Excellence “Severo Ochoa” award to the Instituto de Astrofisica de Canarias. A.M.S. acknowledges support from FCT through the fellowship 2020.05387.BD. M.S. acknowledges financial support from the SNSF for project 200021\_200726.  E.E-B. acknowledges financial support from the European Union and the State Agency of Investigation of the Spanish Ministry of Science and Innovation (MICINN) under the grant PRE2020-093107 of the Pre-Doc Program for the Training of Doctors (FPI-SO) through FSE funds.
\end{acknowledgements}

\bibliographystyle{aa} 
\bibliography{bib.bib} 

\begin{appendix}

\onecolumn

\section{\kpvsys\ plots.}
\label{app_kpvsys}

\begin{figure}[h!]
    \centering
    \includegraphics[width=0.9\hsize]{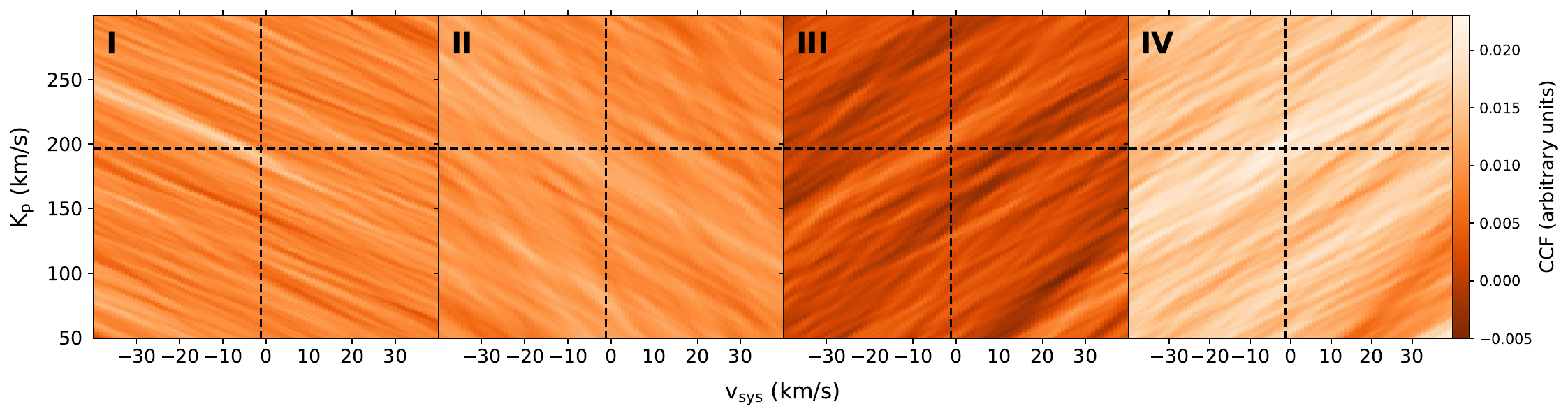}
    \includegraphics[width=0.9\hsize]{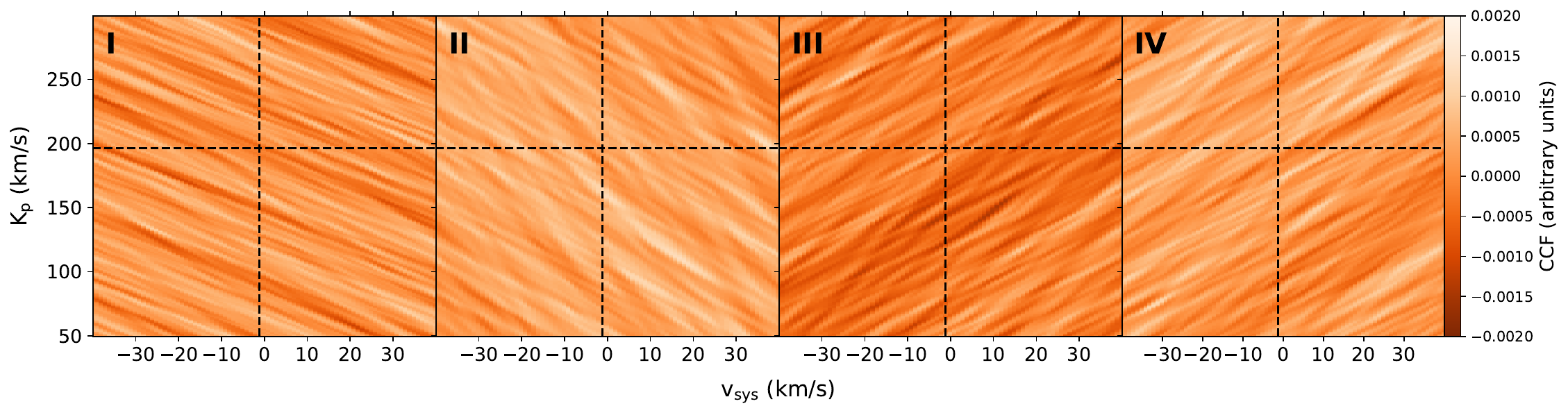}
    \caption{\kpvsys\ plots of \feI\ and \feII\ for each epoch. The dashed black lines mark the expected position of the planetary signal. \textbf{Top:} \feI. The signal is detected in all epochs, though at different amplitudes. \textbf{Bottom:} \feII. No detections are found (see Sect. \ref{sect:feII_discussion} for discussion).}
    \label{fig:kp_vsys_epochs}
\end{figure}

\section{P-T profile exploration}
\label{app_pt_profiles}

\begin{table}[h!]
    \caption{Details of the P-T profiles used to create templates in \texttt{petitRADTRANS}.}
    \label{tab:tp_details}
    \centering
    \begin{tabular}{c c c c c}
        \hline \hline
        \noalign{\smallskip}
        P-T \# & T$_\mathrm{lower}$ (K) & T$_\mathrm{upper}$ (K) & P$_\mathrm{lower}$ (bar) & P$_\mathrm{upper}$ (bar)\\
        \noalign{\smallskip}
        \hline
        \noalign{\smallskip}
        1  & 1700 & 3500  & 0.106136 & 0.001095 \\
        2  & 1700 & 3500  & 5.546040 & 0.057201 \\
        3  & 1700 & 3500  & 0.011466 & 0.000118 \\
        4  & 2200 & 3500  & 0.106136 & 0.001095 \\
        5  & 1700 & 3500  & 0.467905 & 0.004826 \\
        6  & 1700 & 3000  & 0.106136 & 0.001095 \\
        7  & 1700 & 4000  & 0.106136 & 0.001095 \\
        8  & 1200 & 3500  & 0.106136 & 0.001095 \\
        9  & 1700 & 4500  & 0.106136 & 0.001095 \\
        10 & 1700 & 5000  & 0.106136 & 0.001095 \\

        \noalign{\smallskip}
        \hline
    \end{tabular}
    \tablefoot{Pressure (P) and temperature (T) values are given for the lower and upper boundaries of the thermal inversion layer.}
\end{table}

\begin{figure}[h]
    \centering
    \includegraphics[width=0.7\hsize]{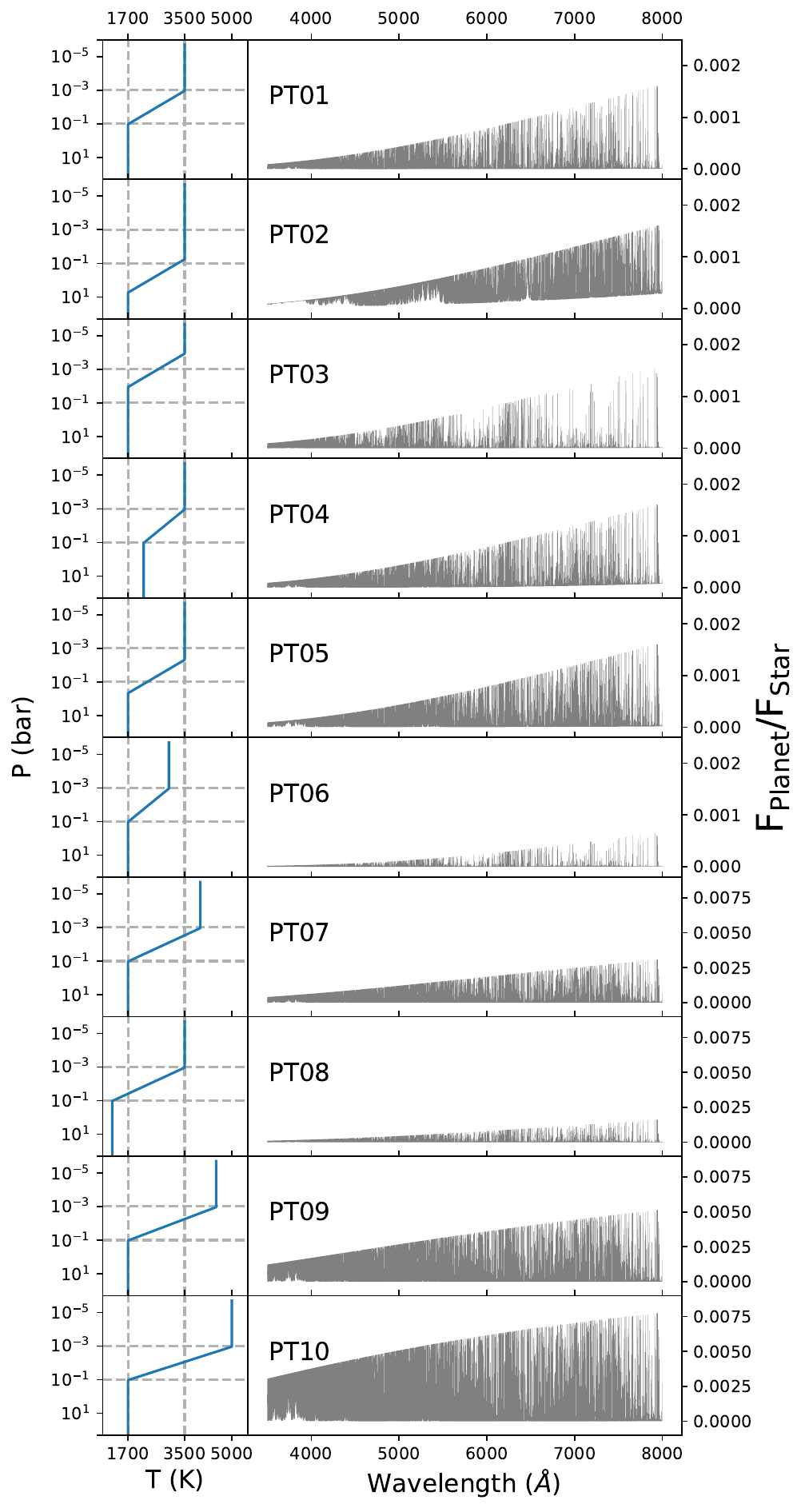}
    \caption{\texttt{petitRADTRANS} templates (right) corresponding to the P-T profiles (left), which were utilised for computing the \feI\ CCFs. We note that the bottom four models are presented on a larger y-scale than the first six. See Table \ref{tab:tp_details} for more details on the atmospheric profiles.}
    \label{fig:prt_models_pt_exploration}
\end{figure}

\end{appendix}

\end{document}